\documentclass[review]{elsarticle}

\usepackage{mathtools}
\usepackage{graphicx}
\usepackage{verbatim}
\usepackage[labelfont=bf]{caption}
\usepackage{subcaption}
\usepackage{siunitx}
\usepackage{booktabs} 
\usepackage{epstopdf}
\usepackage{threeparttable}
\usepackage{longtable}
\usepackage{floatrow}
\usepackage{pgfplots}
\usepackage{pgfplotstable}
\usepackage{amsmath}
\usepackage{url}
\usepackage{indentfirst}
\usepackage{pdfpages}
\usepackage{tikz}
\usepackage{rotating}
\usepackage{xcolor}
\usepackage{float}
\usepackage{threeparttable}
\usepackage{numprint}
\usepackage{rotating}
\usepackage{adjustbox}
\usepackage{calc}

\floatstyle{plaintop}
\restylefloat{table}

\npdecimalsign{.}
\nprounddigits{3}

\journal{Magnetic Resonance Imaging}

\newcommand{\subcptbox}[2]{%
    \makebox[\textwidth/#1 - 1ex][c]{%
        \scriptsize\sffamily\bfseries\relax#2%
    }%
}
\newcommand{\twosubcaptions}[2]{%
    \raisebox{0.75ex}{%
        \centering%
        \subcptbox{2}{#1}\hspace{1ex}%
        \subcptbox{2}{#2}\hspace{1ex}%
    }%
    \vspace*{-0.75ex}
}
\newcommand{\threesubcaptions}[3]{%
    \raisebox{0.75ex}{%
        \centering%
        \subcptbox{3}{#1}\hspace{1ex}%
        \subcptbox{3}{#2}\hspace{1ex}%
        \subcptbox{3}{#3}%
    }%
    \vspace*{-0.75ex}%
}
\newcommand{\foursubcaptions}[4]{%
    \raisebox{0.75ex}{%
        \centering%
        \subcptbox{4}{#1}\hspace{1ex}%
        \subcptbox{4}{#2}\hspace{1ex}%
        \subcptbox{4}{#3}\hspace{1ex}%
        \subcptbox{4}{#4}%
    }%
    \vspace*{-0.75ex}
}

\newcommand{\revised}[1]{\textcolor{black}{#1}}

\begin{document} 

\begin{frontmatter}
    \title{Automatic brain tissue segmentation in fetal MRI using convolutional neural networks}
    
    \author[1]{N. Khalili}
    \author[1]{N. Lessmann}
    \author[2,3]{E. Turk}
    \author[2,3]{N. Claessens}
    \author[4]{R. de Heus}
    \author[2]{T. Kolk}
    \author[1,3,5]{M.A. Viergever}
    \author[2,3]{M.J.N.L. Benders}
    \author[1,3]{I. I\v sgum}
    
    \address[1]{Image Sciences Institute, University Medical Center Utrecht, Utrecht, The Netherlands}
    \address[2]{Department of Neonatology, Wilhelmina Children’s Hospital, University Medical Center Utrecht, Utrecht, The Netherlands}
    \address[3]{Brain Center Rudolf Magnus, University Medical Center Utrecht, Utrecht, The Netherlands}
    \address[4]{Department of Obstetrics, University Medical Center Utrecht, The Netherlands}
    \address[5]{Utrecht University, Utrecht, The Netherlands}

    \begin{abstract}
    MR images of fetuses allow clinicians to detect brain abnormalities in an early \revised{stage} of development. The cornerstone of volumetric and morphologic analysis in fetal MRI is segmentation of the fetal brain into different tissue classes. Manual segmentation is cumbersome and time consuming, hence automatic segmentation could substantially simplify the procedure. However, automatic brain tissue segmentation in these scans is challenging owing to artifacts including intensity inhomogeneity, caused in particular by spontaneous fetal movements during the scan. Unlike methods that estimate the bias field to remove intensity inhomogeneity as a preprocessing step to segmentation, we propose to perform segmentation using a convolutional neural network that exploits images with synthetically introduced intensity inhomogeneity as data augmentation. The method first uses a CNN to extract the intracranial volume. Thereafter, another CNN with the same architecture is employed to segment the extracted volume into seven brain tissue classes: cerebellum, basal ganglia and thalami, ventricular cerebrospinal fluid, white matter, brain stem, cortical gray matter and extracerebral cerebrospinal fluid. To make the method applicable to slices showing intensity inhomogeneity artifacts, the training data was augmented by applying a combination of linear gradients with random offsets and orientations to image slices without artifacts. 
To evaluate the performance of the method, Dice coefficient (DC) and Mean surface distance (MSD) per tissue class were computed between automatic and manual expert annotations. When the training data was enriched by simulated intensity inhomogeneity artifacts, the average achieved DC over all tissue classes and images increased from 0.77 to 0.88, and MSD decreased from 0.78~mm to 0.37~mm. These results demonstrate that the proposed approach can potentially replace or complement preprocessing steps, such as bias field corrections, and thereby improve the segmentation performance.
    \end{abstract}

\end{frontmatter}

\section{Introduction}
\label{sec:intro}  

\begin{figure}[t]
    \centering
        \includegraphics[width=\textwidth]{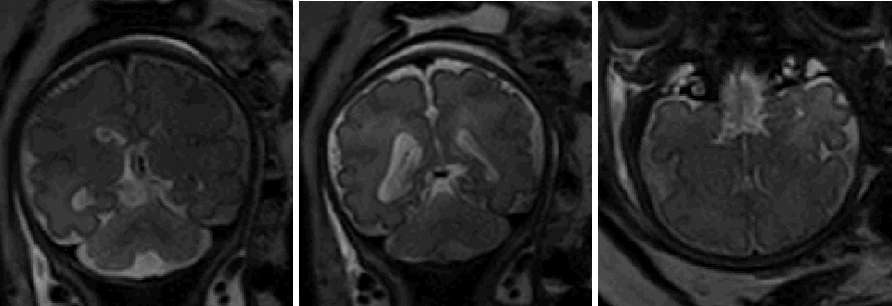}

    \caption{Examples of slices with intensity inhomogeneity in T2 weighted MRI of the fetus cropped to the brain. Note that the full slices also visualize a larger part of the \revised{fetus} as well as the maternal body.}
    \label{fig:example_slices}
\end{figure}

Important neurodevelopmental changes occur in the last trimester of pregnancy, i.e., between 30 and 40 weeks of gestation, including volumetric growth, myelination and cortical gyrification \cite{kostovic2006development,zilles2013development,schaer2012measure}. Magnetic resonance imaging (MRI) is widely used to non-invasively assess and monitor the developmental status of the fetal brain \emph{in utero} \cite{twickler2002fetal,levine2003fast}. The cornerstone of volumetric and morphologic analysis in fetal MRI is the segmentation of the fetal brain into different tissue classes, such as white and gray matter. Performing this segmentation manually, however, is extremely time-consuming and requires a high level of expertise. The reasons are not only the complex convoluted shapes of the different tissues, but also the limited image quality due to imaging artifacts. Fetal MR imaging is particularly challenging in this regard because the \revised{receiver} coils can only be positioned on the maternal body and not closer to the anatomy of interest. Furthermore, movements of the fetus relative to the mother can only to some extent be controlled and predicted. Especially\revised{,} fetal motion therefore negatively affects the image quality and causes artifacts such as intensity inhomogeneity (Figure~\ref{fig:example_slices}).

\begin{figure}[t]
    \includegraphics[width=\textwidth]{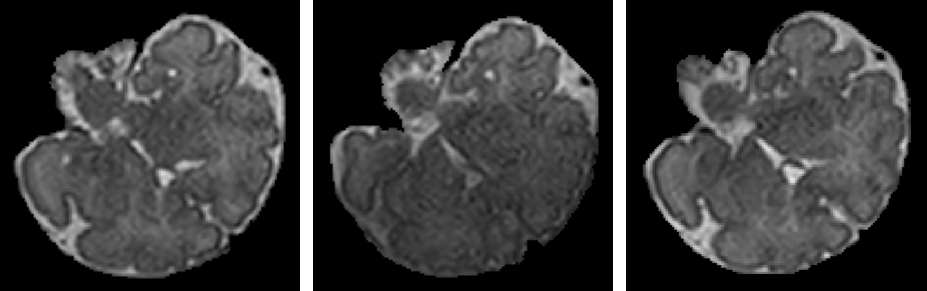}
    \threesubcaptions{Slice N\,$-$\,1}{Slice N}{Slice N\,$+$\,1}
    \caption{Example of a fetal T2 weighted MRI with intensity inhomogeneity (middle) and the slice before (left) and after (right) from the same scan, both without artifacts. Structures outside the fetal cranium have been masked out.}
    \label{fig:example_slices_before_after}
\end{figure}

Because manual annotation is very time consuming and additionally hampered by these artifacts, a reliable automatic tissue segmentation tool would provide a valuable alternative, especially if it could give detailed fetal brain tissue segmentations in the presence of artifacts. To cope with imaging artifacts, previous approaches in the literature performed the segmentation in images reconstructed from multiple 2D acquisitions. Most fetal MRI scans are acquired in 2D using single-shot fast spin-echo (SSFSE) sequences \cite{yamashita1997mr}. Artifacts such as intensity inhomogeneity may therefore appear only in some slices, e.g., due to movements during acquisition of these slices, but do not have to be present in their immediate neighboring slices as well (Figure~\ref{fig:example_slices_before_after}). Volumetric reconstruction approaches are typically based on the acquisition of several stacks of 2D slices in axial, sagittal and coronal orientation. These stacks are registered to a common coordinate space so that they can be combined into a single reconstructed 3D volume, thus removing artifacts that affect only some slices and inter-slice inconsistencies \cite{jiang2007mri,gholipour2010robust,kuklisova2012reconstruction,kainz2015fast, ebner2018automated}.

For automatic segmentation of fetal brain tissue in reconstructed MR volumes, Habas et al.\ \cite{habas2010atlas} proposed a method using an atlas-based \revised{expectation maximization (EM)} model to segment white matter (WM), gray matter, germinal matrix, and \revised{extracerebral cerebrospinal fluid (eCSF)}. Prior to performing the segmentation, another EM model was used for bias field correction.
Gholipour et al.\ \cite{gholipour2012multi} proposed a method for segmentation of the ventricles in fetal MRI. As a preprocessing step, in addition to using volumetric reconstructions, intensity inhomogeneity was corrected using the non-parametric entropy maximization method \cite{gholipour2011fetal}. Initial segmentation was obtained with the use of STAPLE \cite{warfield2004simultaneous}, then the final segmentation was derived with a probabilistic shape model that incorporates intensity and local spatial information.
Serag et al.\ \cite{serag2012multi} proposed an atlas-based brain segmentation method for both neonatal and fetal MRI. Fetal scans were reconstructed into a single 3D brain volume using the slice-to-volume reconstruction method described in \cite{jiang2007mri} and intensity inhomogeneity was removed using the N4 algorithm \cite{tustison2010n4itk}. \revised{Thereafter, the} fetal brain scans were segmented into cortex, ventricles and hemispheres. 



\revised{Deep learning methods have recently been very successful and have often outperformed traditional machine learning and model-based methods in medical image analysis \cite{litjens2017survey} including brain MRI \cite{akkus2017review,makropoulos2018review}. A major strength of these networks is their ability to extract the features relevant for the tasks directly from the data.} There is no need anymore to first derive a set of handcrafted features from the image as input to a classifier or model, the networks rather learn themselves to extract and interpret features relevant to the segmentation task. \revised{Therefore, deep learning methods often achieve a better performance than traditional machine learning methods with hand-crafted features.} \revised{However, CNNs usually require large sets of diverse training data. To enlarge the size of training set and to ensure robustness to expected variability in the data, some studies use data augmentation techniques such as random rotation, random translation and random noise injection \cite{zhang2018abnormal,wang2018alcoholism}}. We therefore hypothesize that, while artifacts such as intensity inhomogeneity are challenging for traditional approaches and therefore normally require preprocessing of the images, CNNs may be able to adapt and become invariant to such artifacts if they are presented enough examples during training. However, manual segmentation of slices with intensity inhomogeneity is much more cumbersome than segmentation of artifact free slices so that a sizable training database is difficult to obtain. We therefore propose to tackle one of the most common artifacts in fetal MRI, namely intensity inhomogeneity, by randomly adding synthetic intensity inhomogeneity to slices for which a corresponding reference segmentation is available. By only mutating the intensity values but not the orientation or shape of structures in the image, the same reference segmentation can be used as ground truth. This tailored data augmentation strategy affects network training only. At inference time, in contrast to previous methods, no complex preprocessing of the image is required.



Furthermore, previous methods focused on segmenting the brain into the three main tissue classes: WM, cortical gray matter and ventricles. However, characteristics of other tissue classes, such as cerebellum (CB) and brain stem (BS), are important to understand and predict healthy or aberrant brain development in preterm infants of similar gestational age as fetuses \cite{bouyssi2016third}. The cerebellum is particularly of clinical interest as it is one of the fastest growing brain regions during the last trimester of pregnancy \cite{dobbing1973quantitative}. 


Another challenge for segmentation of the fetal brain in MRI is the large field of view of these scans. Since the fetus is scanned \emph{in utero}, the images also visualize parts of the maternal and the fetal body, and not only the head of the fetus as would be the case in regular brain MRI. Similar to previous publications \cite{ivsgum2015evaluation, moeskops2015automatic}, we therefore propose to first automatically segment the intracranial volume (ICV) of the fetus to identify the region of interest. A number of studies proposed segmentation of the ICV in fetal MRI \cite{anquez2009automatic,rajchl2017deepcut,nadieh2017,salehi2017real}. 
Following our previous work \cite{nadieh2017}, we segment the ICV directly in the entire image to fully automatically detect a region of interest.


The method we propose performs segmentation of fetal and brain tissues. The method first identifies the ICV from the fetal MRI slices using a convolutional neural network. Subsequently, the identified volume is segmented by another 2D convolutional neural network. \revised{Note that the proposed approach is applied to 2D slices of images reconstructed in a standard way, i.e. without reconstruction to high resolution volumes. The contribution of this paper is twofold: First, we propose data augmentation technique that synthesizes intensity inhomogeneity artifacts to improve the robustness against these artifacts. Second, the fetal brain segmentation is performed into seven classes: CB, basal ganglia and thalami (BGT), ventricular cerebrospinal fluid (vCSF), WM, BS, cortical gray matter (cGM) and eCSF in contrast to previous methods which focused on WM, cGM and cerebrospinal fluid only.}

The remainder of this paper is organized as follows: in \revised{Section} 2 the data set used for the method development and evaluation is described, in \revised{Section} 3 the method for fetal brain segmentation and the simulation of intensity inhomogeneity are described, in \revised{Section} 4 the evaluation method is given. The performed experiments and their results are presented in \revised{Section} 5, followed by a discussion of the method and the results in \revised{Section} 6. Our conclusions are given in the final section.

\section{Data}
\label{sec:data}

\subsection{Fetal MRI dataset}

This study includes T2-weighted MR scans of 12 fetuses (22.9--34.6 weeks post menstrual age). Images were acquired on a Philips Achieva 3T scanner at the University Medical Center (UMC) Utrecht, the Netherlands, using a turbo fast spin-echo sequence. Repetition time (TR) was set to \revised{2793} ms, echo time (TE) was set to 180 ms and the flip angle to 110 degrees. 
The acquired voxel size was  $1.25\times1.25\times2.5$ mm$^3$, the reconstructed voxel size was $0.7\times0.7\times1.25$ mm$^3$, and the reconstruction matrix was $512\times512\times80$. \revised{The images were not reconstructed using high resolution framework. }\revised{We included images acquired in coronal plane as they appeared less affected by fetal motion in comparison with images acquired in axial or sagittal plane.} The orientation of the fetal brain was determined using fast survey scanning. During the scan, the mother was lying on her left side to reduce the chance of inferior vena cava syndrome. \revised{The local ethical board approved the study and parental informed consent was obtained}.

\begin{figure}[t!]
    \includegraphics[width=\textwidth]{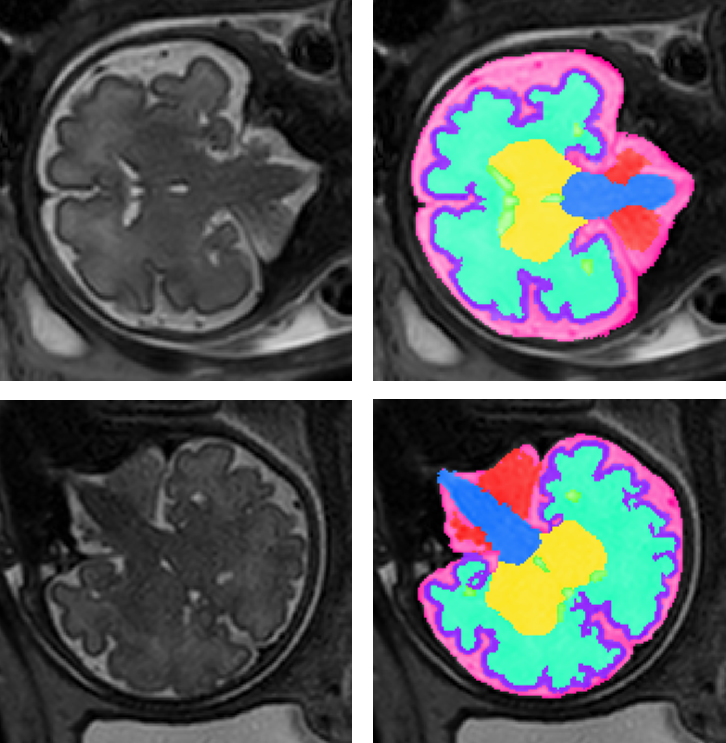}
    \twosubcaptions{Image slice}{Manual annotation}

    \includegraphics[width=\textwidth]{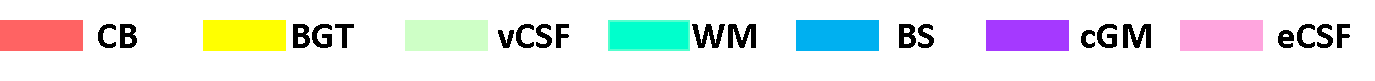}

    \caption{Examples of manual reference segmentation in coronal fetal MRI, showing slices cropped to the region of interest (first column) and the slices overlaid with the manual segmentations (second column).}
    \label{fig:manual_annotation}
\end{figure}

The reference standard was defined by manual annotation of all scans in 2D slices by a trained medical student. The brain was segmented into seven tissue classes: CB, BGT, vCSF, WM, BS, cGM and eCSF. Annotation was accomplished by manual pixel-wise painting of the brain tissues in each coronal image slice using an in-house developed software. The labeling of each of the seven classes was indicated by a color overlay (Figure~\ref{fig:manual_annotation}). The software allowed the user to zoom-in, zoom-out and scroll through the slices during the manual segmentation. The manual segmentation protocol was identical to the protocol described by I\v sgum et al.\ \cite{ivsgum2015evaluation} for neonatal brain tissue segmentation. The ICV was defined as the union of all manually segmented tissue classes.

In total 15 slices (1.5\%), 7 in the training set and 8 in the test set, were  too distorted by severe motion artifacts to be manually annotated. In total 126 of the remaining, manually annotated slices (26.2\%), 32 in the training set and 94 in the test set, were identified as affected by intensity inhomogeneity that hampered manual annotation.

\subsection{Neonatal MRI dataset}
While we propose a method aimed specifically at segmentation of fetal MRI scans, the proposed segmentation approach and especially the data augmentation technique that simulates intensity inhomogeneity (detailed in the following section) might be useful for brain segmentations in MRI scans. Therefore, to evaluate the applicability of this technique to a different MRI data set, we additionally included nine brain MR scans of preterm born infants.

T2-weighted MR images were acquired on a Philips Achieva 3T scanner at the University Medical Center Utrecht, the Netherlands. The images were made with Turbo Field Echo (TFE) and Turbo Spin Echo (TSE) sequences with TR set to 4847 ms and TE set to 150 ms. The scans were acquired at 40 weeks of post menstrual age in the coronal plane. The acquired voxel size was  $ 0.35\times0.35\times1.2$ mm$^3$ and the reconstruction matrix was $512\times512\times110$. The reference standard was defined by manual annotation of an expert into seven tissue types (CB, BGT, vCSF, WM, BS, cGM and eCSF) \revised{in MR scans without visible intensity inhomogeneity artifacts}. Manual annotation was performed using the same protocol as described above for fetal MRI. These scans are part of the NeoBrainS12 segmentation challenge \cite{NeoBrainS2} and did not show intensity inhomogeneity artifacts.

\revised{The remaining four scans showed intensity inhomogeneity and they were used as a test set}. However, manual reference segmentations in these scans were not available and, consequently, we evaluated the segmentation performance in this set only by visual inspection.

\section{Method}
\begin{figure}[ht]
    \includegraphics[width=\textwidth]{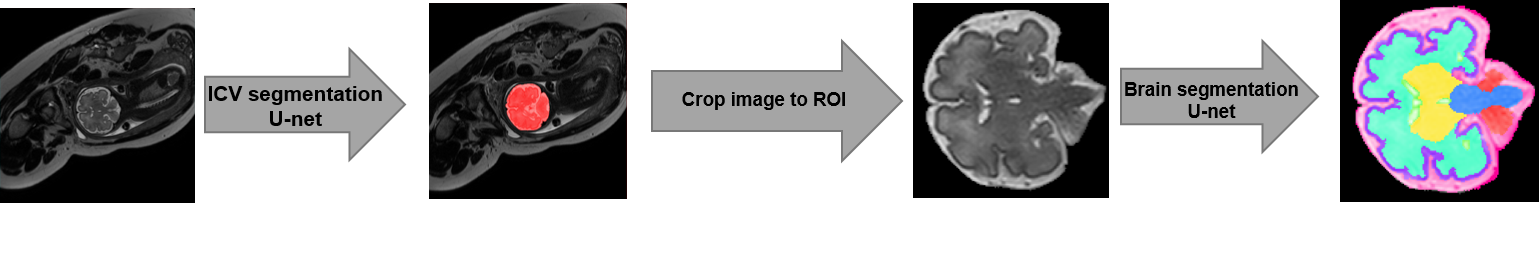}
    \includegraphics[width=\textwidth]{colorcode_figure.png}
    \caption{Proposed pipeline for automatic tissue segmentation method. The fetal ICV is automatically segmented in the original slice. \revised{Thereafter, the image is automatically cropped to the region of interest (ROI) so that the} tissue segmentation can be restricted to the fetal ICV only.}
    \label{fig:pipeline}
\end{figure}

To simplify the brain tissue segmentation and allow the segmentation method to focus on the fetal brain only, the fetal ICV is first automatically extracted. Subsequently, the identified ICV is automatically segmented into seven tissue classes. An overview of this pipeline is shown in Figure~\ref{fig:pipeline}. \revised{The same network architecture, described in Section 3.1, was used for ICV extraction and brain tissue segmentation.}

\subsection{Brain segmentation} 

\begin{figure}[ht]
    \includegraphics[width=\textwidth]{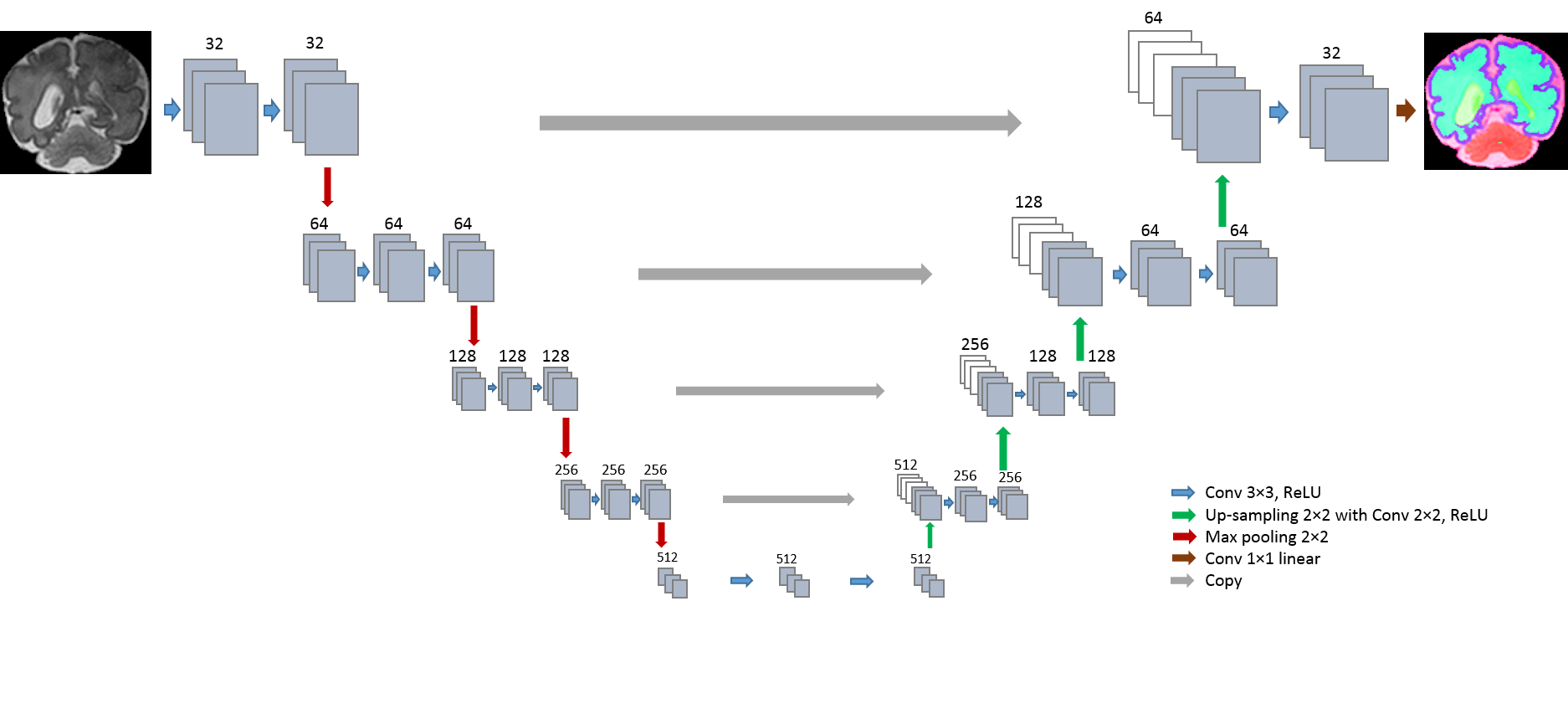}
    \caption{Network architecture: The network consists of a contraction path and an expansion path. The contraction path consists of repeated convolution layers followed by max pooling, and the expansion path consists of convolution layers followed by upsampling.}
    \label{fig:network}
\end{figure}

Extraction of ICV and its subsequent segmentation into seven brain tissue classes are achieved with two fully convolutional networks (FCN) with identical U-net architecture \cite{ronneberger2015u} trained with 2D slices. Each network is trained independently to perform its specific task. The U-net architecture consists of a contracting path and an expanding path. The contracting path consists of repeated $3\times3$ convolutions followed by rectified linear units (ReLUs). A $2\times2$ max pooling downsamples the features. The number of feature channels doubles after every two convolutional layers. In the expansion path, an up-sampling is followed by a $2\times2$ convolution which halves the number of feature channels. The results are concatenated with the corresponding contraction path and convolved by two $3\times3$ convolutional layers followed by a ReLU. At the final layer, one $1\times1$ convolutional layer maps each component of the feature vector to the desired number of classes. Batch normalization \cite{ioffe2015batch} is applied after all convolutional layers to allow for faster convergence.  The network architecture is illustrated in Figure~\ref{fig:network}.

\revised{Both networks were trained using stochastic gradient descent with back propagation.} We optimized both networks using the Adam optimizer with Nesterov momentum \cite{kingma2015method,dozat2016incorporating} using a fixed learning rate of 0.0001. Standard data augmentation techniques, namely random flipping and rotation, were used during training to increase the variation of the training data and to mimic different orientations of the fetal brain. The slices were flipped in horizontal and vertical direction with 50\% probability and were rotated with a rotation angle randomly chosen between 0 and 360 degrees. 

To first identify the intracranial area \revised{in the image}, a U-net was trained by minimizing the cross-entropy between network output and manual segmentation for all pixels in each slice in each training batch. The convolutional neural network was trained with batches of 12 slices in each iteration. Given that the network performs voxel classification, ICV segmentation may result in small isolated clusters of false positive voxels outside the ICV. These were removed by discarding 3D connected components smaller than 3 cm$^3$. \revised{This threshold was empirically chosen in this study. It was chosen large enough to remove small false positive clusters of voxels, and also small enough to prevent removing any parts of the brain that are not fully connected in 3D. The latter is often the case in scans with substantial motion artifacts where the signal is lost in one slice and consequently, the ICV segmentation is not fully connected in 3D.}


The segmented intracranial fetal areas were further segmented into seven tissue classes using another\revised{, separately trained,} U-net. Each pixel in the image was classified as either CB, BGT, vCSF, UWM, BS, cGM, eCSF or background. In contrast to the network for ICV segmentation, this network was trained by maximizing the Dice coefficient between network output and manual segmentation. This was done to achieve robustness against an imbalance of samples from the different classes. This network was trained with batches of 18 slices in each iteration. \revised{We have implemented the network in Keras, an open-source neural-network library written in Python \cite{chollet2015keras}.} 

\subsection{Intensity inhomogeneity augmentation (IIA)}
\label{sec:method_iia}

\begin{figure}[ht!]
\centering
\resizebox{\textwidth}{!}{%
    \includegraphics[clip,scale = 0.73]{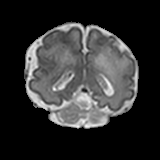}
    \includegraphics[clip,scale = 0.73]{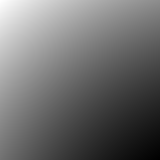}
    \includegraphics[clip,scale = 0.73]{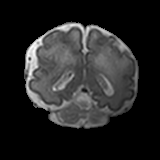}
}\\[0.8ex]
\resizebox{\textwidth}{!}{%
    \includegraphics[clip,scale = 0.73]{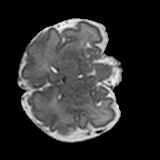}
    \includegraphics[clip,scale = 0.73]{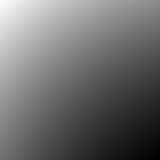}
    \includegraphics[clip,scale = 0.73]{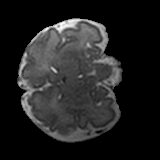}
}\\[0.8ex]
\resizebox{\textwidth}{!}{%
    \includegraphics[clip,scale = 0.73]{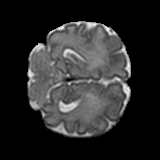}
    \includegraphics[clip,scale = 0.73]{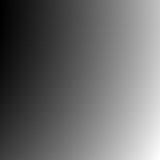}
    \includegraphics[clip,scale = 0.73]{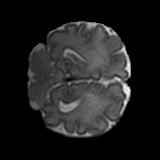}
}\\[0.8ex]
\resizebox{\textwidth}{!}{%
    \includegraphics[clip,scale = 0.73]{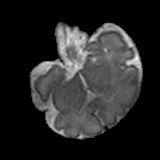}
    \includegraphics[clip,scale=0.73]{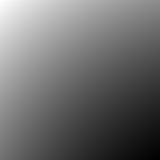}
    \includegraphics[clip,scale = 0.73]{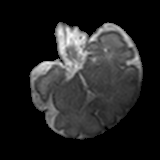}
}
\threesubcaptions{Artifact-free slice}{Random inhomogeneity pattern}{Slice with synthetic inhomogeneity}
\caption{Example of coronal slices with simulated intensity inhomogeneity. Original slice (first column), simulated intensity inhomogeneity pattern (second column) and the image slice after adding the synthetic intensity inhomogeneity artifact (third column).}
\label{fig:simulate}
\end{figure}

To make the network segmenting brain tissue classes robust to intensity inhomogeneity artifacts, we trained this network with slices containing simulated intensity inhomogeneity artifacts. The artifacts were simulated by applying a combination of linear gradients with random offsets and orientations to \revised{a slice without intensity inhomogeneity artifacts (I)}:
\begin{equation}\label{eq:1}
Z = I\times((X+x_0)^{2}+(Y+y_0)^{2}),
\end{equation}
where X and Y are 2D matrices with integer values from zero to the size of the image in x and y direction, respectively. The offsets $x_0$ and $y_0$ control the balance between the x and y components and were randomly chosen from different ranges ($x_0$: [43, 187]; $y_0$: [-371, 170]). The optimal ranges were found with a random hyperparameter search. Additionally, the gradient patterns were randomly rotated between 0 to 360 degree to mimic intensity inhomogeneity in various directions. These random components, offsets and rotation, result in inhomogeneity patterns that allow the network to become invariant to the location and orientation of regions with low and decreasing contrast. The intensities in both the original slices as well as the slices with simulated intensity inhomogeneity were normalized to the range $[0, 1023]$ before feeding them to the network. 
Figure~\ref{fig:simulate} shows examples of two slices from a fetal MRI scan with added synthetic intensity inhomogeneity.

\section{Evaluation}

The automatic brain tissue segmentation was evaluated by means of the Dice coefficient (DC) for volume overlap and the mean surface distance (MSD) between manual reference segmentation and automatically obtained segmentation. In the fetal MRI scans, these metrics were calculated in 2D, i.e., per slice, and were then averaged across all slices. In the neonatal MRI scans, following previous work \cite{moeskops2015automatic}, these metrics were calculated in 3D.

\section{Experiments and Results}

In our experiments, we first evaluated the overall segmentation performance of the proposed pipeline with respect to the different tissue classes. \revised{To evaluate the influence of the proposed intensity inhomogeneity augmentation technique with the standard augmentation techniques, the segmentation performance before and after applying intensity inhomogeneity augmentation was evaluated.} Furthermore, we evaluated whether this augmentation technique is able to generalize to different data, i.e., whether it leads to similar performance improvements in neonatal brain segmentation.
The fetal MRI dataset was randomly divided into a training and a test set. Each set contained 6 scans. The neonatal MRI dataset was divided into a training set with 3 scans and a test set with the remaining \revised{6} scans for which the manual reference segmentations \revised{of two scans} were available. \revised{The training and test sets are listed in Table~\ref{tab:TrainingTest}.}

\begin{table}[]
\resizebox{\textwidth}{!}{

\begin{tabular}{l|l|l}

& Training set& Training set description                                                          \\ \hline
Set 1 (fetal)     & 6 scans &Excluding slices with intensity inhomogeneity artifacts\\ \hline
Set 2 (fetal)    & 6 scans &Including slices with intensity inhomogeneity artifacts    \\ \hline
Set 3 (neonatal) & 3 scans &Intensity inhomogeneity not visible in any slice 
\caption{\revised{The fetal MRI dataset was randomly divided into a training and a test set each containing 6 scans. We defined two training sets with fetal data: Set 1 contained 6 scans, but without those slices in which intensity inhomogeneity was visible. Set 2 contained the entire slices of 6 scans. The neonatal MRI dataset was divided into a training set with three scans and a test set with the six remaining scans (Set 3). The test set in all three sets contained 6 scans.}}
\label{tab:TrainingTest}

\end{tabular}
}
\end{table}
\subsection{Segmentation performance}
\begin{figure}[t!]
    \centering
    \resizebox{\textwidth}{!}{%
        \includegraphics[trim={15cm 4cm 16cm 4cm},clip]{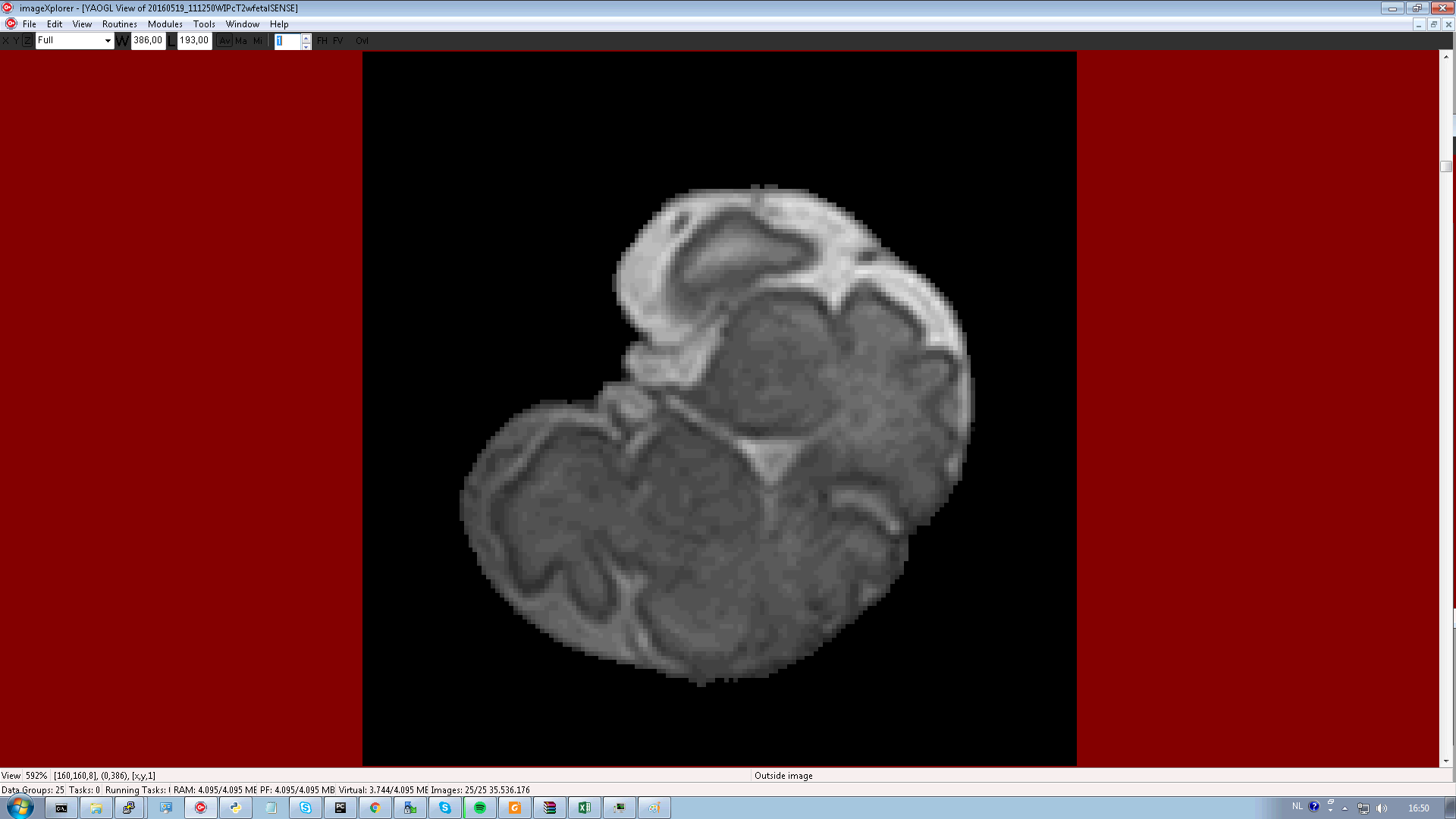}%
        \hspace{6ex}%
        \includegraphics[trim={15cm 4cm 16cm 4cm},clip]{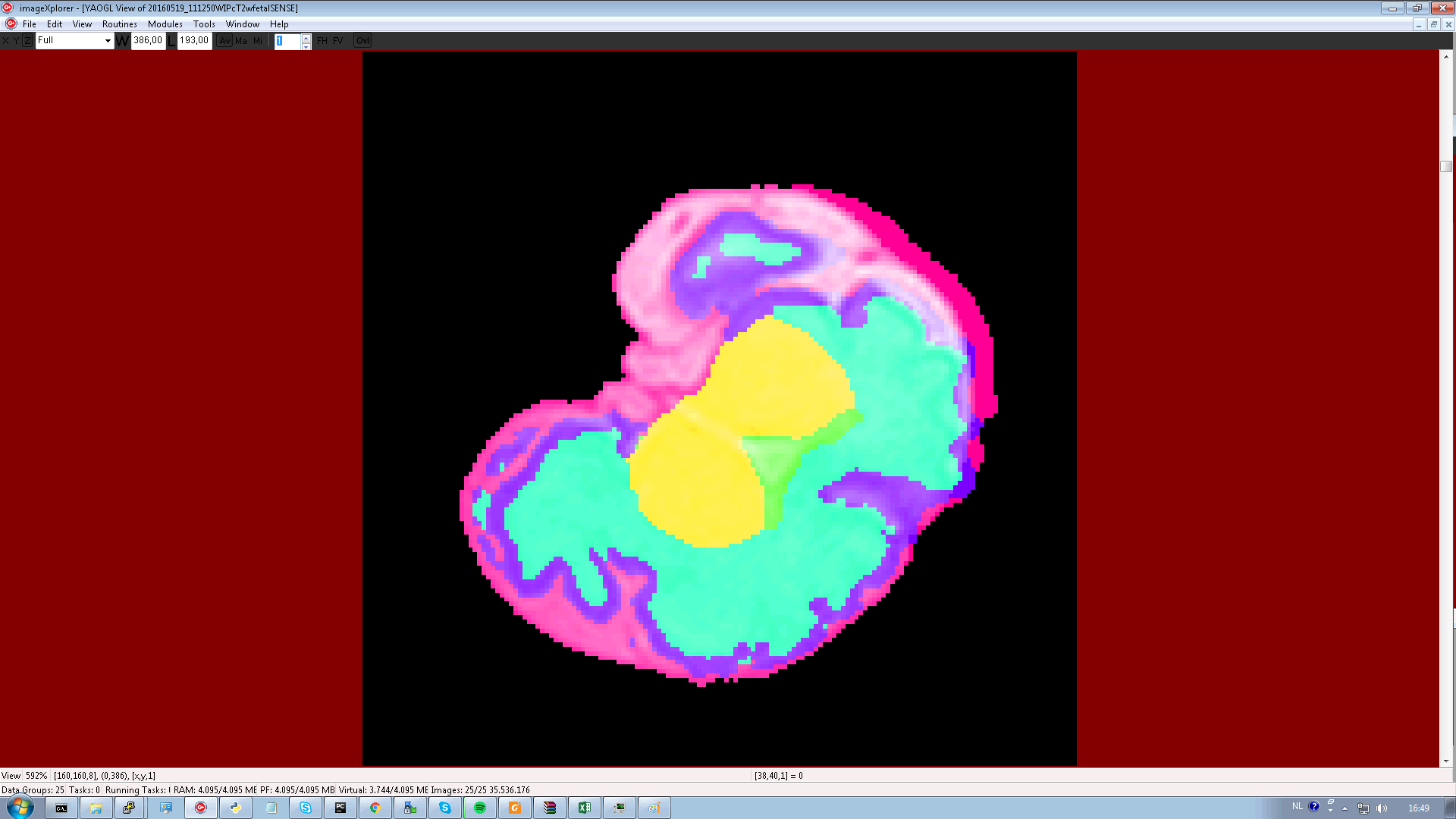}%
        \hspace{6ex}%
        \includegraphics[trim={15cm 4cm 16cm 4cm},clip]{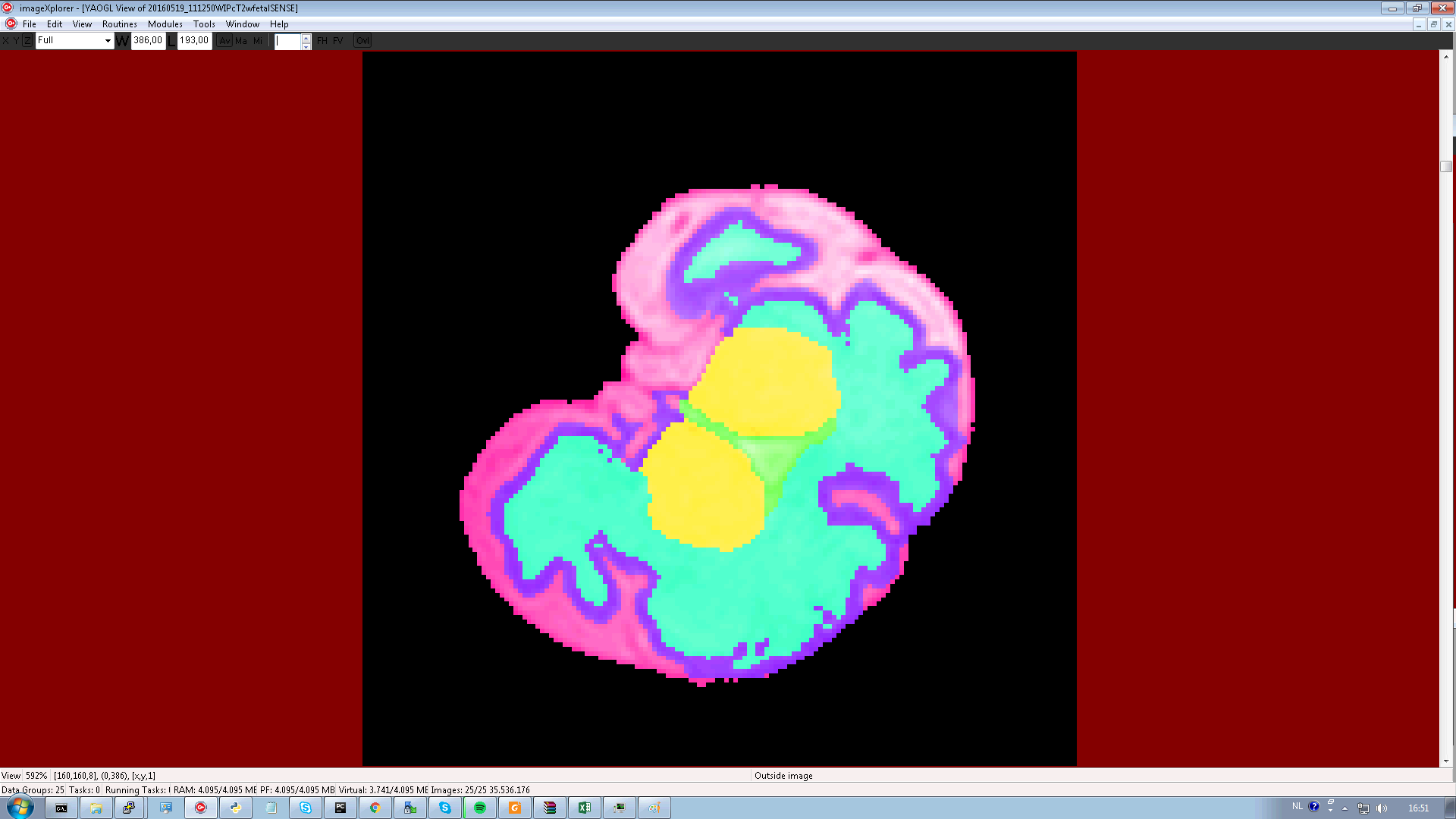}%
        \hspace{6ex}%
        \includegraphics[trim={15cm 4cm 16cm 4cm},clip]{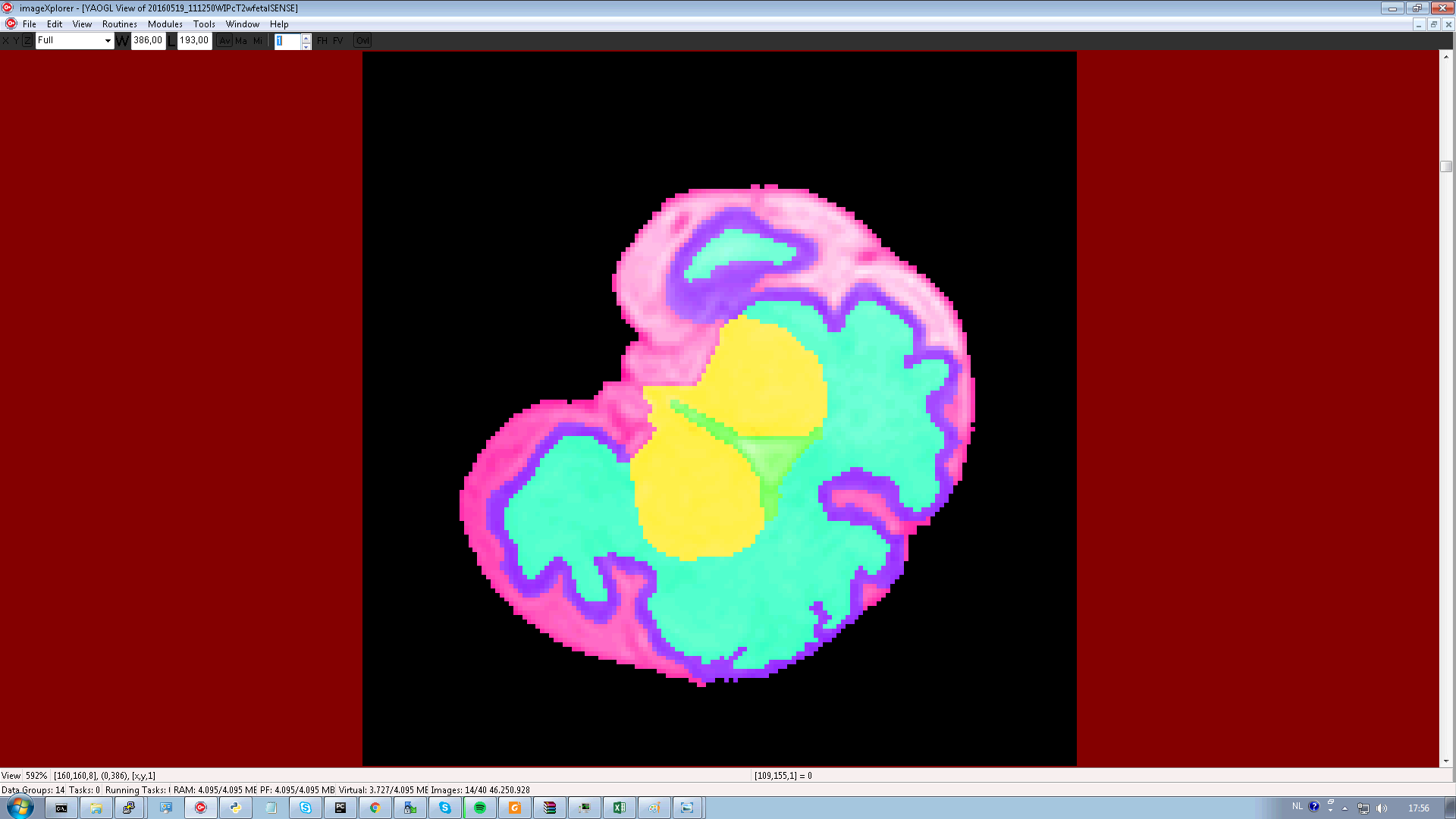}%
    }\\[0.8ex]
    \resizebox{\textwidth}{!}{%
        \includegraphics[trim={15cm 4cm 16cm 4cm},clip]{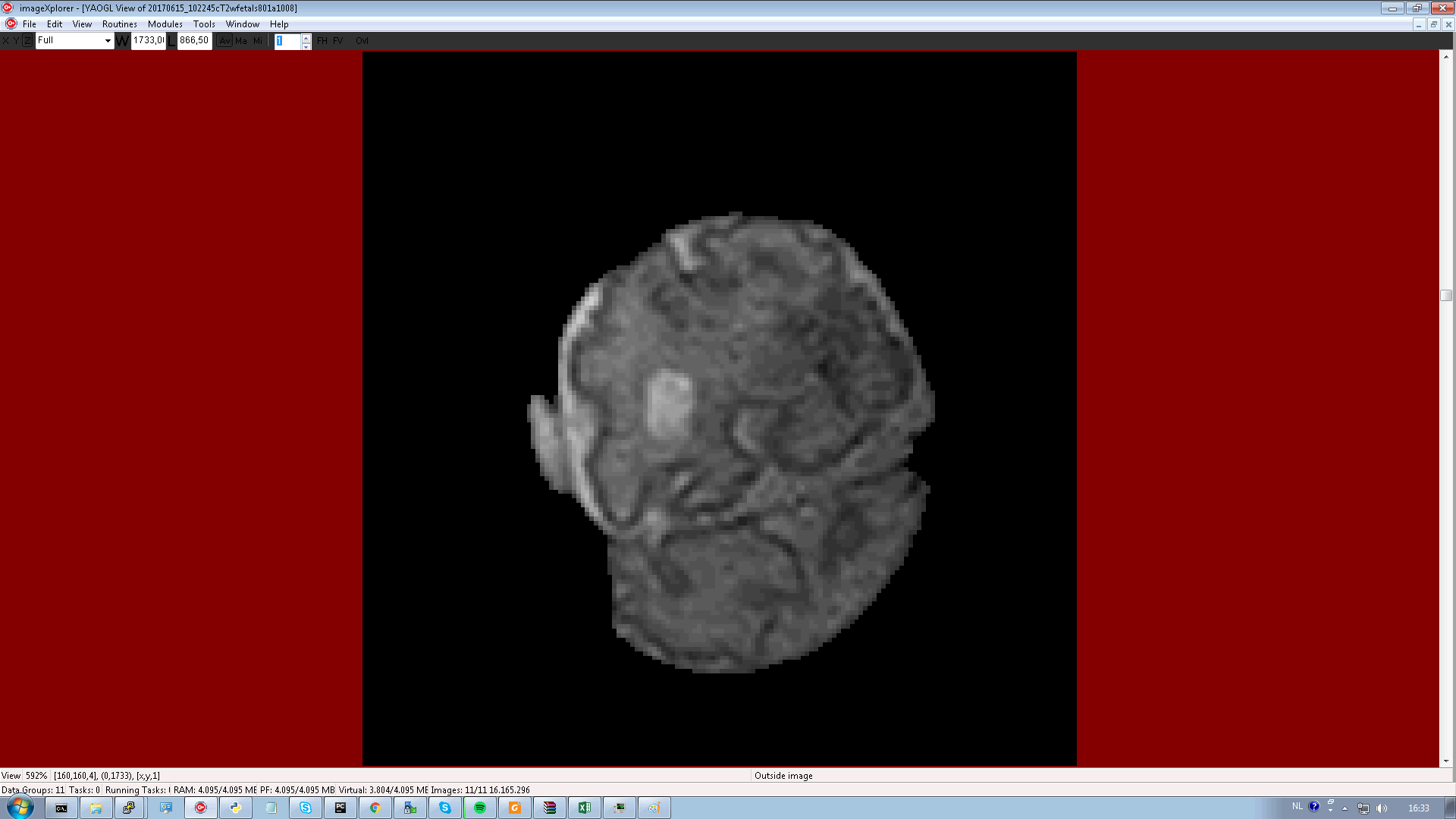}%
        \hspace{6ex}%
        \includegraphics[trim={15cm 4cm 16cm 4cm},clip]{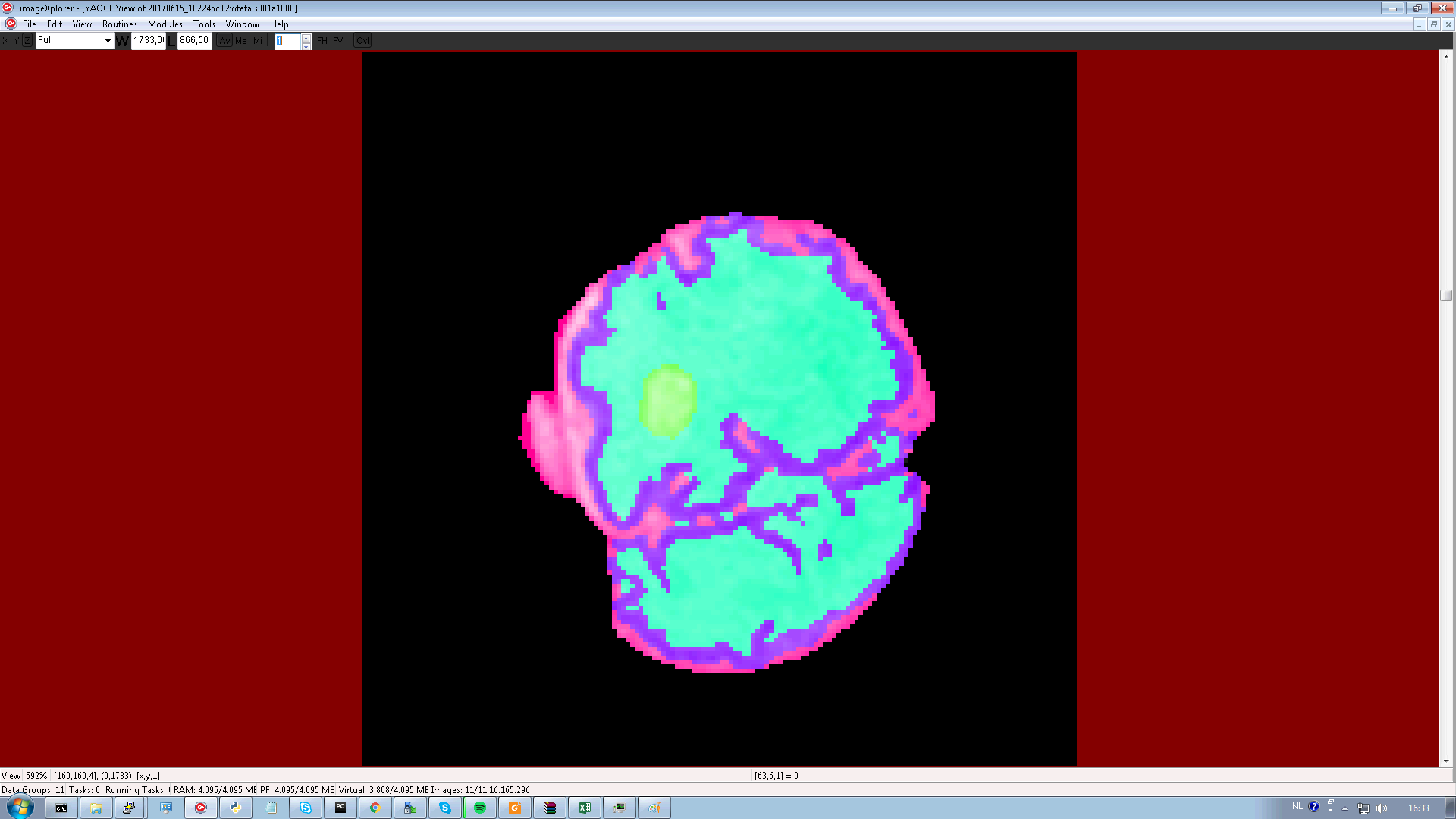}%
        \hspace{6ex}%
        \includegraphics[trim={15cm 4cm 16cm 4cm},clip]{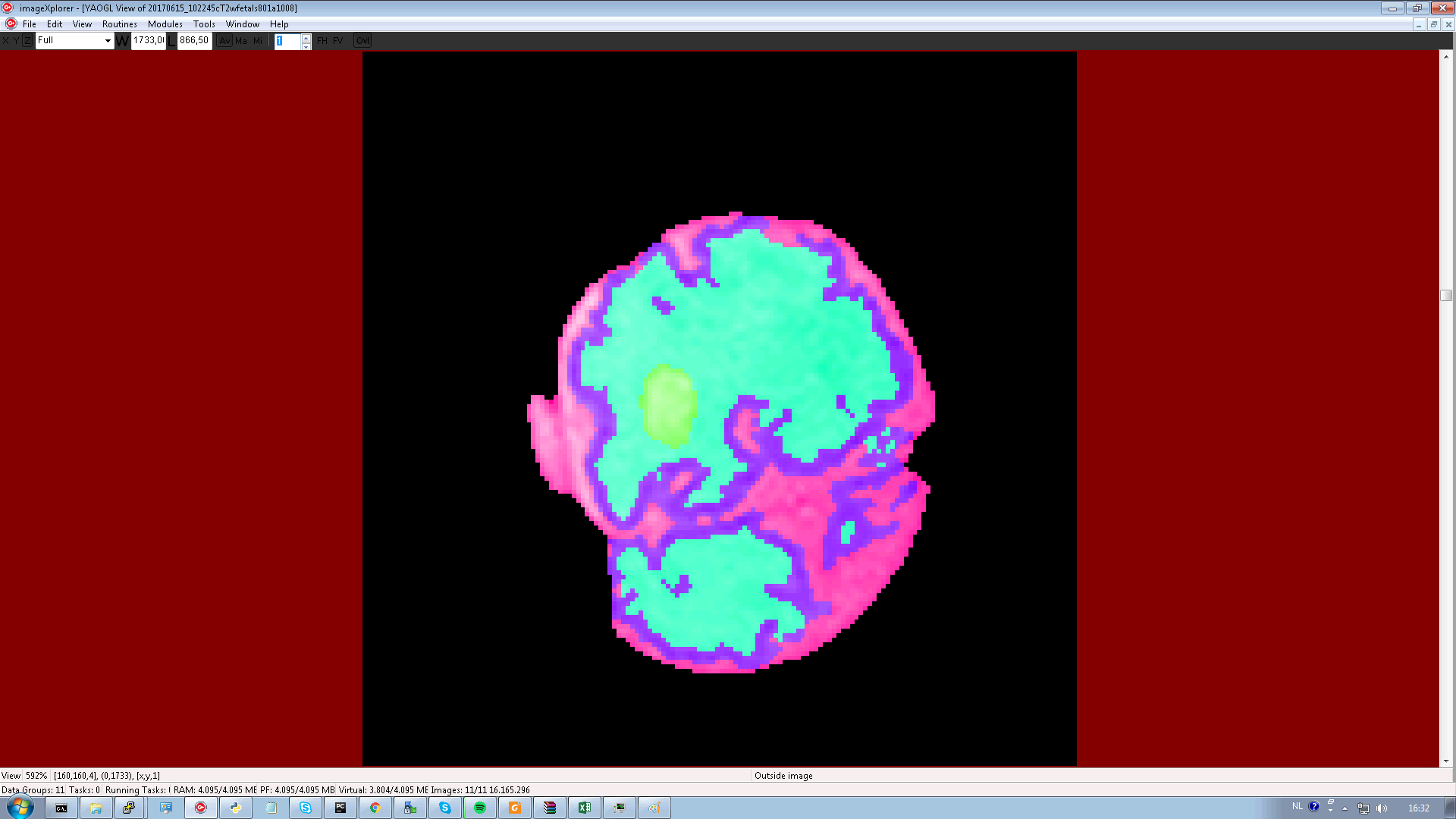}%
        \hspace{6ex}%
        \includegraphics[trim={15cm 4cm 16cm 4cm},clip]{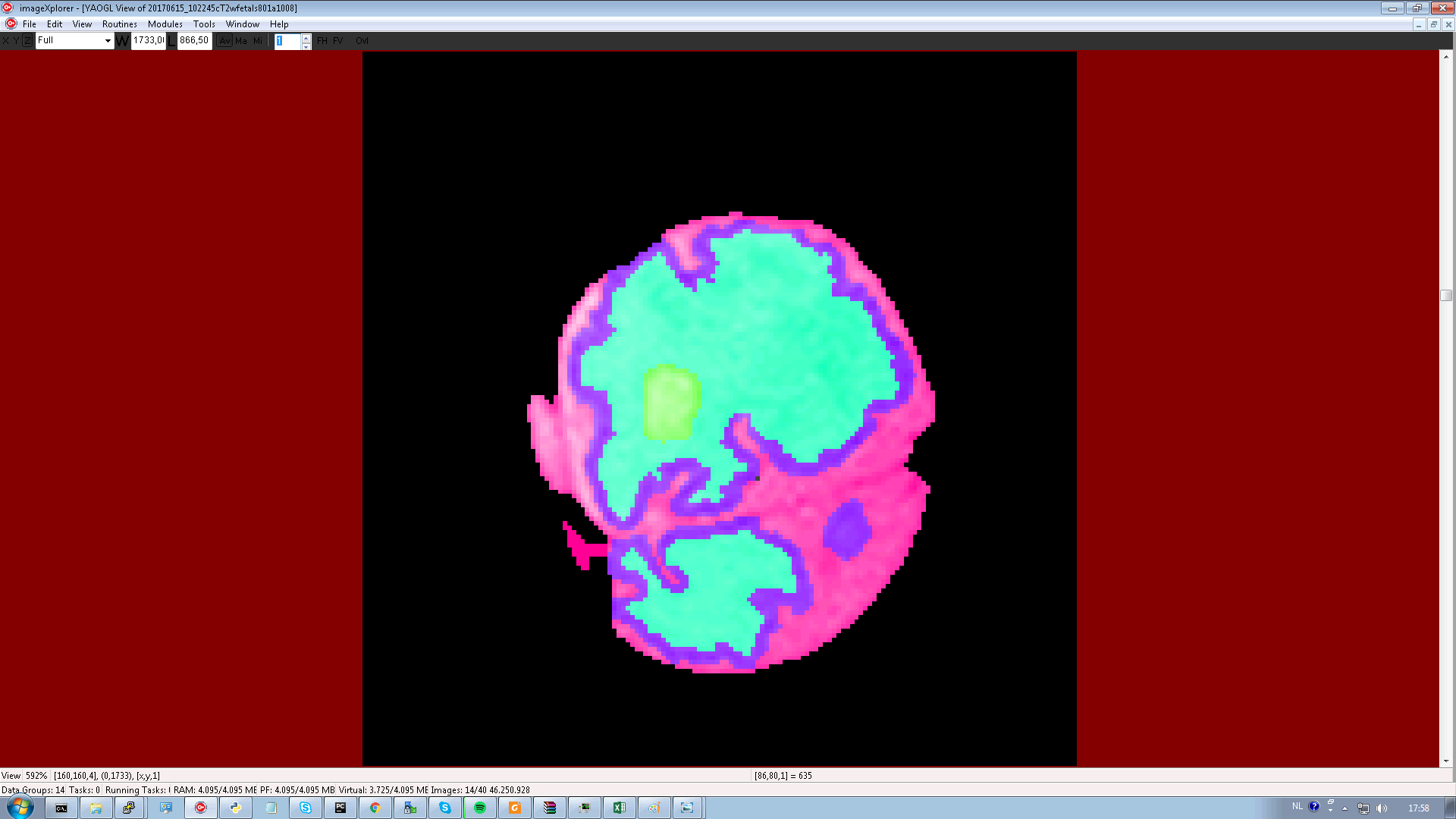}%
    }\\[0.8ex]
    \resizebox{\textwidth}{!}{%
        \includegraphics[trim={15cm 4cm 16cm 4cm},clip]{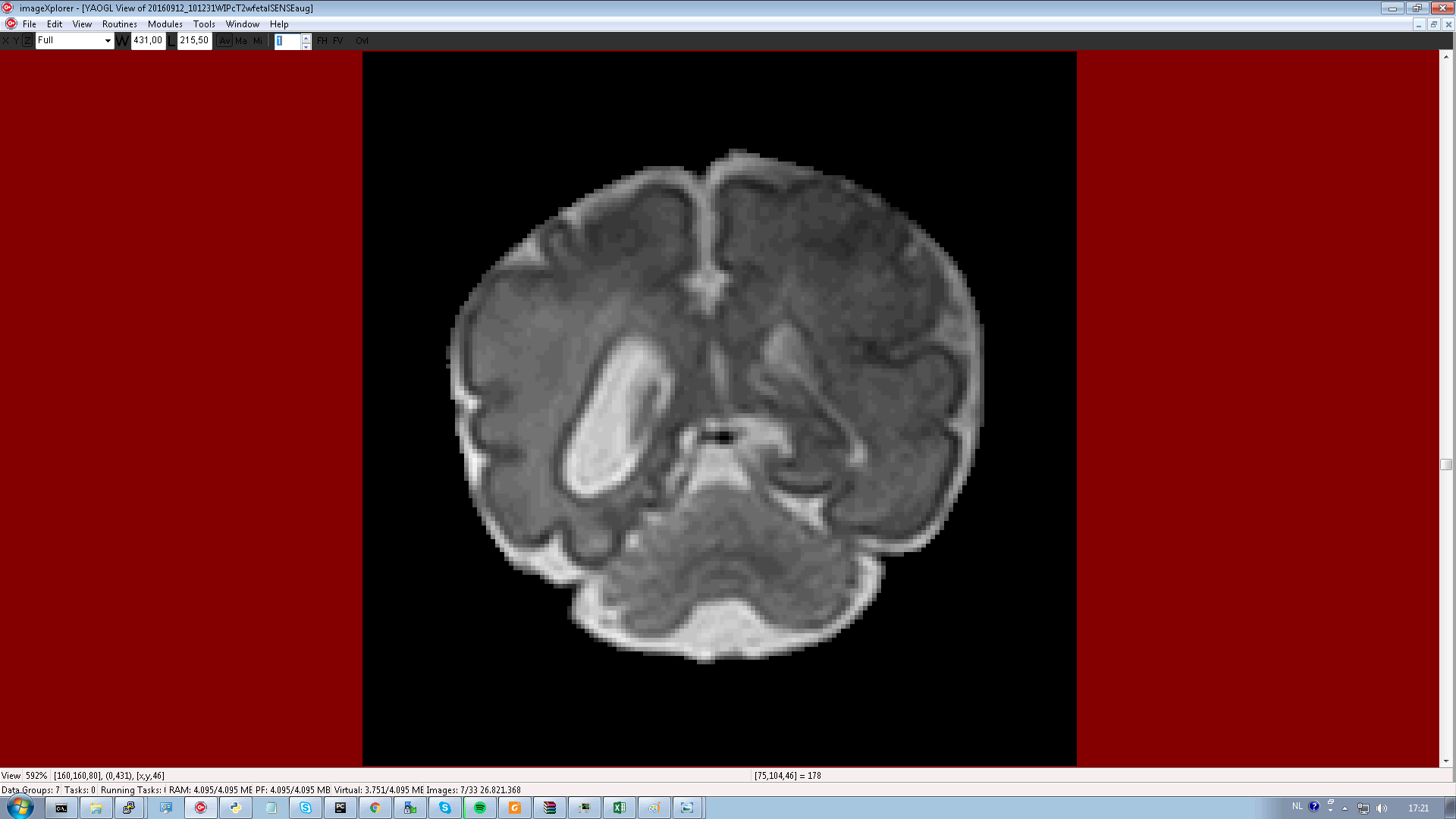}%
        \hspace{6ex}%
        \includegraphics[trim={15cm 4cm 16cm 4cm},clip]{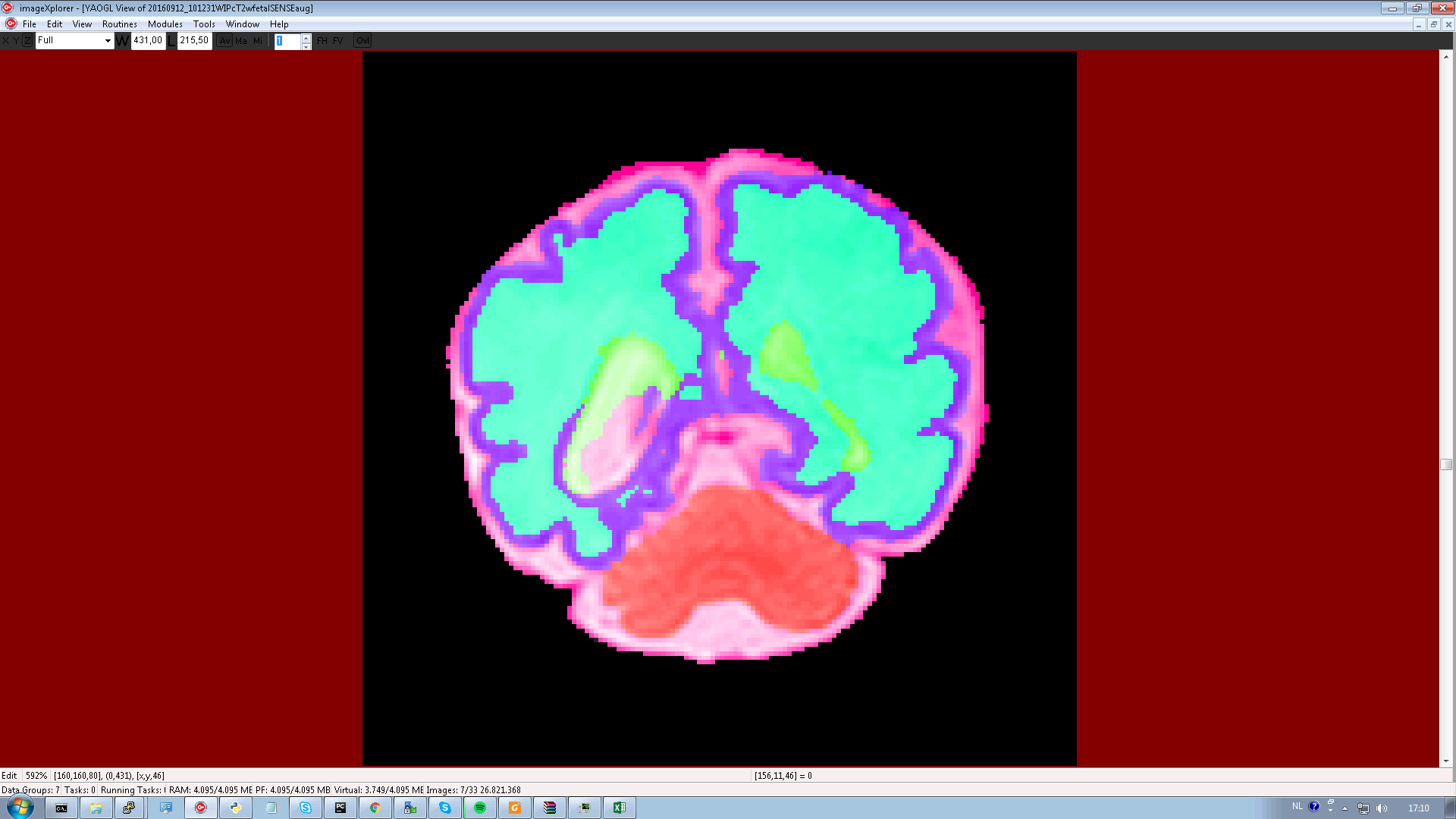}%
        \hspace{6ex}%
        \includegraphics[trim={15cm 4cm 16cm 4cm},clip]{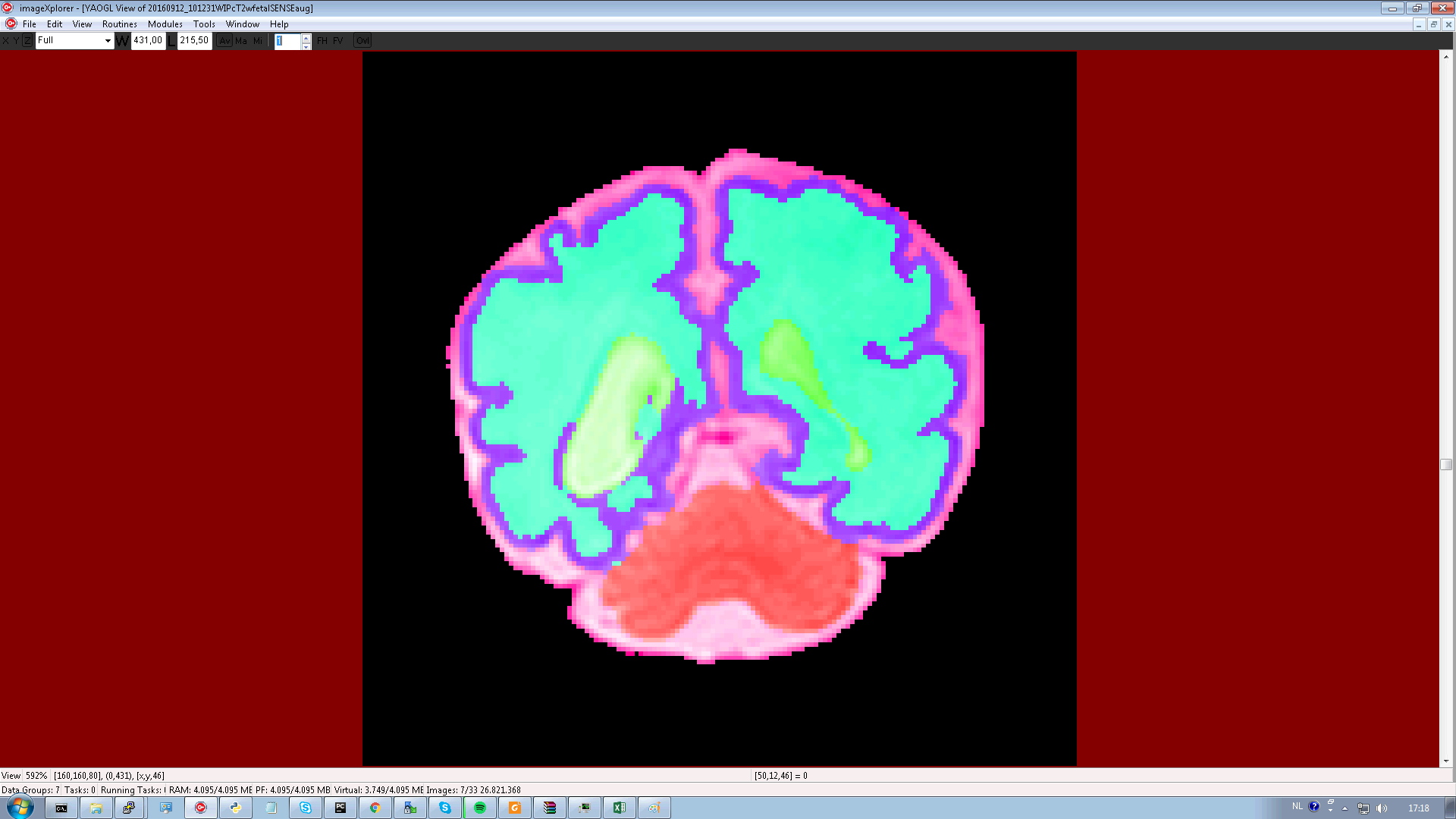}%
        \hspace{6ex}%
        \includegraphics[trim={15cm 4cm 16cm 4cm},clip]{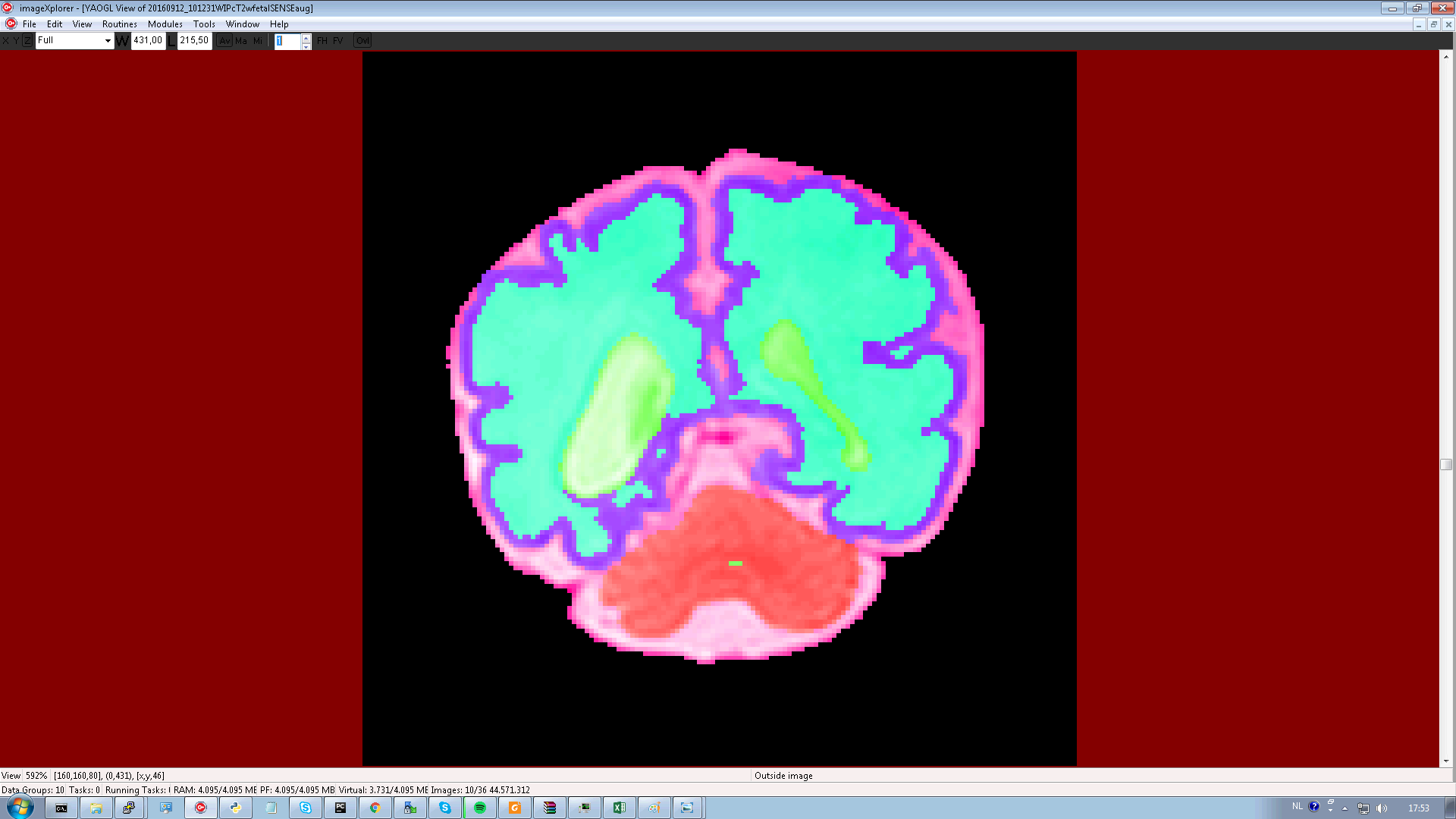}%
    }
    \foursubcaptions{Image slice}{Flip \& Rotation}{Flip \& Rotation \& IIA}{Reference}
        \includegraphics[width=\textwidth]{colorcode_figure.png}

    \caption{Examples of automatic brain tissue segmentation in slices with visible intensity inhomogeneity. A slice from T2-weighted fetal MRI scan with visible intensity inhomogeneity (first column); segmentation obtained with network only using flipping and rotation augmentation (second column); segmentation obtained with network using IIA (third column); manual reference segmentation (fourth column).}
    \label{fig:inhomogeneity}
\end{figure}

\begin{figure}[t!]
    \centering
    \resizebox{\textwidth}{!}{%
        \includegraphics[width=\textwidth]{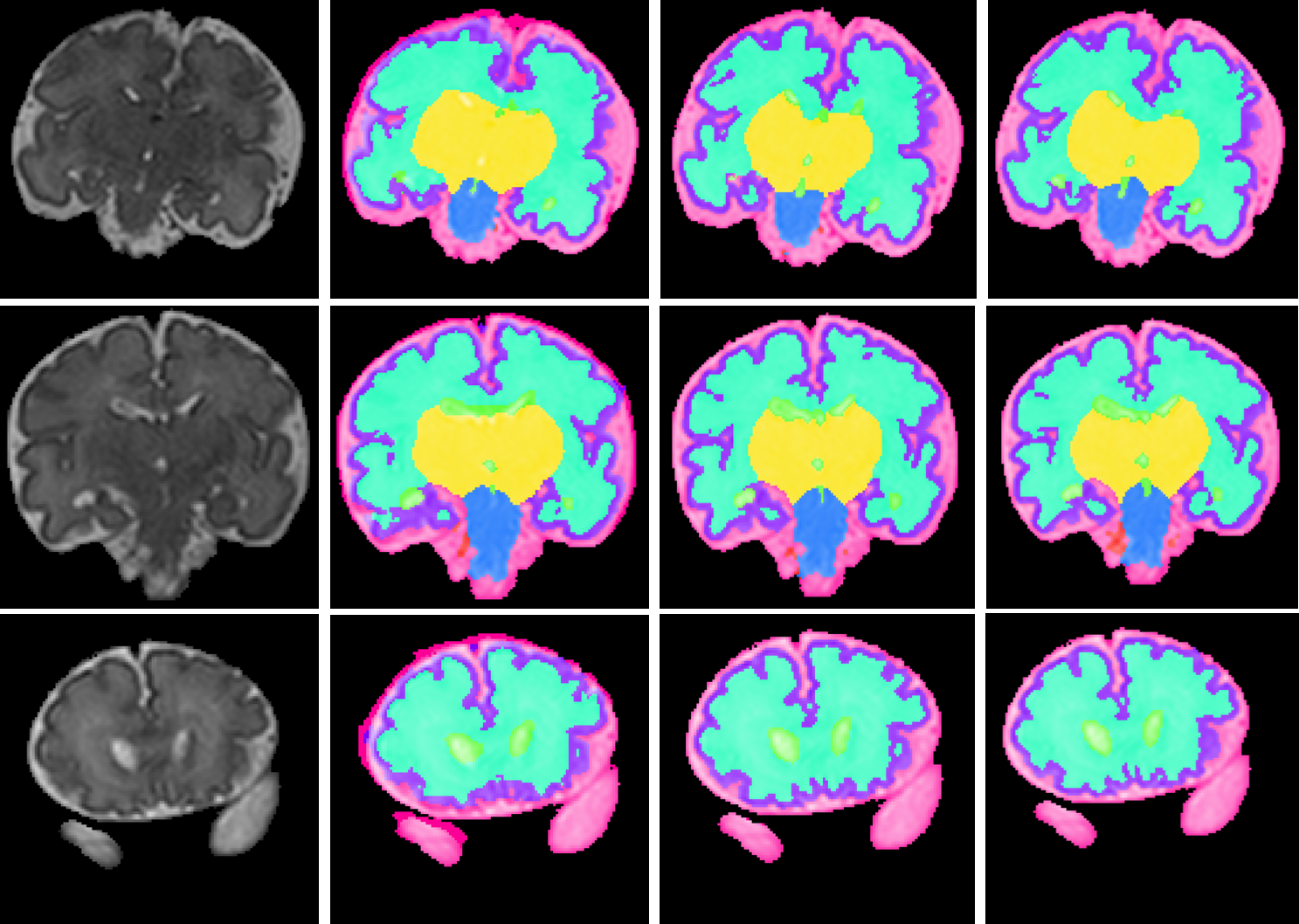}%

    }\\[0.8ex]

    \foursubcaptions{Image slice}{Flip \& Rotation}{Flip \& Rotation \& IIA}{Reference}
    \includegraphics[width=\textwidth]{colorcode_figure.png}

    \caption{Examples of automatic brain tissue segmentation in fetal images acquired in the slices without visible intensity inhomogeneity. A slice from T2-weighted fetal MRI scan with visible intensity inhomogeneity (first column); segmentation obtained with network using only flipping and rotation augmentation (second column); segmentation obtained with network additionally using IIA (third column); manual reference segmentation (fourth column).}
    \label{fig:normal}
\end{figure}

The performance of the proposed method using standard augmentation and the proposed IIA as described in Section 3 was evaluated. Slices with intensity inhomogeneity artifacts resulting from image acquisition were excluded from the training data. The average performance in the six test scans is listed in Table~\ref{tab:exclude_II} for each of the seven tissue classes \revised{(Set 1)}. The average DC ranged from 0.80 for CB to 0.94 for eCSF and the average MSD ranged from 0.62~mm for CB to 0.18~mm for BS. Furthermore, to evaluate whether IIA improves the performance differently in slices with artifacts than in artifact free slices, we compared the performance on slices with clearly visible intensity inhomogeneity artifacts and slices without visible artifacts. These results are also listed in Table \ref{tab:exclude_II}. As shown, the automatic segmentations were less accurate on slices with intensity inhomogeneity strong enough to hamper manual annotation compared with slices without visible intensity inhomogeneity. Figure~\ref{fig:inhomogeneity} and Figure~\ref{fig:normal} illustrate the segmentation performance on slices with intensity inhomogeneity and without visible intensity inhomogeneity, respectively.
Note that these results were obtained with networks trained without any slices with intensity inhomogeneity artifacts resulting from the image acquisition, but slices with simulated intensity inhomogeneity were used for training. Excluding slices with intensity inhomogeneity from the training set is a more realistic training scenario because manual segmentation of such slices is cumbersome and reference segmentations for such slices might therefore not be available. However, extending the training set with image slices affected by real intensity inhomogeneity artifacts but in which manual annotation was still feasible could potentially further improve the performance. We therefore trained the networks also including slices with intensity inhomogeneity artifacts resulting from image acquisition \revised{(Set 2)}. The quantitative results are listed in Table~\ref{tab:II_include}. As shown in the table, in this experiment, we also separately evaluated the performance and impact of IIA in slices with visible and without visible intensity inhomogeneity artifacts.

Finally, we compared the performance of the proposed fetal brain tissue segmentation method with the performance of previous methods (Table~\ref{tab:compare}). The performance of the proposed method was comparable to the performance of other methods, even though it performs a finer segmentation into seven tissue classes instead of only four \cite{habas2010atlas} or three \cite{serag2012multi} tissues. The performance of previous methods is taken from the literature. Hence, the methods have been evaluated using different data set and thus this comparison can provide an indication only.

\begin{table}
\resizebox{\textwidth}{!}{

\begin{tabular}{c|c|c|n{1}{3}n{1}{1}n{1}{1}n{1}{1}n{2}{1}n{1}{1}n{1}{1}|n{1}{1}n{1}{1}}

\npdecimalsign{.}
\nprounddigits{2}
 &&& {CB} & {BGT} & {vCSF} & {WM} & {BS} & {cGM} & {eCSF} & {Mean} \\ \hline
All test slices&IIA&DC  &0,801554227&	0,888975163&	0,875236633	&0,922388436	&0,929966857	&0,829320887&	0,943158672& 0,884371554
\\
&&MSD & 0,619908237&	0,413569634&	0,470044553&	0,383897423&	0,181288454	&0,318488402&	0,187718458&0,367845023
\\ 
& Without IIA &DC  &0,688070178	&0,807221819	&0,724210077&	0,849426329&	0,850392047&	0,672127132&	0,819976685&0,773060609
\\
&&MSD &0,99526716&	0,726272298	&1,331176727&	0,875260677	&0,252849273&	0,768573889	&0,548505887
&0,78541513
\\
\hline
Slices with II&IIA&DC  &0,694278293	&0,901178718	&0,80700113&	0,898969591&	0,946603227	&0,781824955&	0,876718179&	0,843796299
\\
&&MSD & 0,812928808&	0,371959593	&0,807071063&	0,565232868&	0,286212765	&0,41836791	&0,415873687	&0,525378099
\\& Without IIA &DC  &0,467200332	&0,572233279&	0,48461581	&0,743831367&	0,703703419&	0,491712942&	0,678227293& 0,591646349
\\
&&MSD &1,723500976&	1,340236751&	3,742232506	&1,656148232	&0,628969257&	1,275311244	&0,99989491& 1,623756268
\\
\hline
Slices without II&IIA&DC  &0,754404698	&0,913842784	&0,854464057&	0,918010715	&0,926447837	&0,828649917	&0,923117502&0,87413393
\\
&&MSD & 0,428218063	&0,40648691&	0,729010673&	0,426300279	&0,159073699	&0,351260758&	0,241383681&0,391676295
\\
& Without IIA &DC  &0,719168433&	0,836644273&	0,802491099	&0,902208028	&0,933013467&	0,75792872&	0,870225856&	0,831668554
\\&&MSD &0,535858917&	0,502477131&	0,675285696&	0,594564195&	0,207525961&0,446359157&	0,403377463&	0,48077836
\\
\end{tabular}
}
\caption{Performance of fetal brain tissue segmentation into seven tissue classes when the network is trained on slices without intensity inhomogeneity resulting from image acquisition. The network is evaluated on the entire test set and additionally on only the slices with intensity inhomogeneity artifacts (94 slices) and on the slices without visible intensity inhomogeneity artifacts (357 slices). The segmentation performance with intensity inhomogeneity augmentation (IIA) used during training is compared with the performance of the same network without IIA. The results are expressed as the mean Dice coefficient (DC) and the mean surface distance (MSD) in mm.}
\label{tab:exclude_II}
\end{table}

\begin{table}
\resizebox{\textwidth}{!}{

\begin{tabular}{c|c|c|n{1}{3}n{1}{1}n{1}{1}n{1}{1}n{1}{3}n{1}{3}n{1}{1}|n{1}{1}n{1}{1}}

&&& {CB} & {BGT} & {vCSF} & {WM} & {BS} & {cGM} & {eCSF} & {Mean}\\ \hline

All test slices&IIA&DC &0,793820412	&0,931386794	&0,874061628&	0,919396137&	0,946435558	&0,834999148	&0,944196498	&0,892042311
\\
 &&MSD &0,714536276&	0,427694335&	0,433826804	&0,380944219	&0,195135628	&0,307360427	&0,181813539	&0,377330175
\\
&Without IIA &DC &0,77753066&	0,887270158	&0,851371495&	0,923394436&	0,931048782&	0,82075224&	0,939754778&	0,87587465
\\
&&MSD & 0,770822563	&0,501395099	&0,521309188	&0,439886085	&0,192868294&	0,355776017&	0,195266706&	0,425331993
 \\
\hline
Slices with II&IIA&DC &	0,719362	&0,905807748	&0,815572439	&0,901368523	&0,934094878	&0,789953168	&0,877560951&	0,84910281
\\
&&MSD &0,735490541	&0,426395471&	0,702515596&	0,613075346&	0,278362029&	0,39766594&	0,412902729	&0,509486807
\\
&Without IIA &DC & 0,722044897	&0,876948795&	0,842499458	&0,905265569&	0,941064561&	0,801250225	&0,885026782&	0,853442898
\\
&&MSD &0,951250677&	0,534915512	&0,704914067	&0,5206622&	0,266943754	&0,351485856	&0,396958795	&0,532447266
\\
\hline
Slices without II&IIA&DC &0,749368128	&0,905431581&	0,884074744&	0,927142318	&0,926497368&	0,856024372&	0,933508381	&0,883149556
\\
&&MSD &
0,489135842	&0,437670016&	0,388090849&	0,342529916	&0,150107929&	0,272020225	&0,212397513	&0,327421756
\\
&Without IIA &DC & 0,730801753&	0,922798338&	0,888425758	&0,93322419	&0,932840493	&0,85024207&	0,932475281	&0,884401126
\\
&&MSD &	0,520145052&	0,401776248	&0,362919259	&0,36014806	&0,140559923	&0,289740638&	0,214560266&0,32712135
\end{tabular}
}
\caption{Performance of fetal brain tissue segmentation into seven tissue classes when the network is trained with slices with intensity inhomogeneity resulting from the image acquisition. The network is evaluated on the entire test set and additionally on only slices with intensity inhomogeneity (94 slices) and on the slices without intensity inhomogeneity (357). The segmentation performance with intensity inhomogeneity augmentation (IIA) in the training is compared with the performance of the same network without IIA in the training. The results are expressed as the mean Dice coefficient (DC) and the mean surface distance (MSD) in mm.}
\label{tab:II_include}

\end{table}

\subsection{Comparison of data augmentation techniques}

To evaluate the influence of the proposed IIA as well as the influence of the standard data augmentation techniques used in this study (random flipping and random rotation), the following experiments using fetal scans were performed. \revised{In this experiment using fetal scans, all slices with intensity inhomogeneity were removed from the training set (Set 1)}. First, a network was trained without any data augmentation to serve as baseline for the comparison. Second, a network was trained using random flipping of the training slices to augment the training data. Third, a network was trained using random flipping and random rotation of the training slices. Finally, as presented in Section 5.1, a network was trained using random flipping and random rotation, and additionally with IIA, i.e., randomly simulated intensity inhomogeneity. In this last experiment, all slices were manipulated with IIA during the training. Results of this last experiment are listed in Table \ref{tab:exclude_II}.

\begin{figure}[ht!]

    \includegraphics[width=0.85\textwidth]{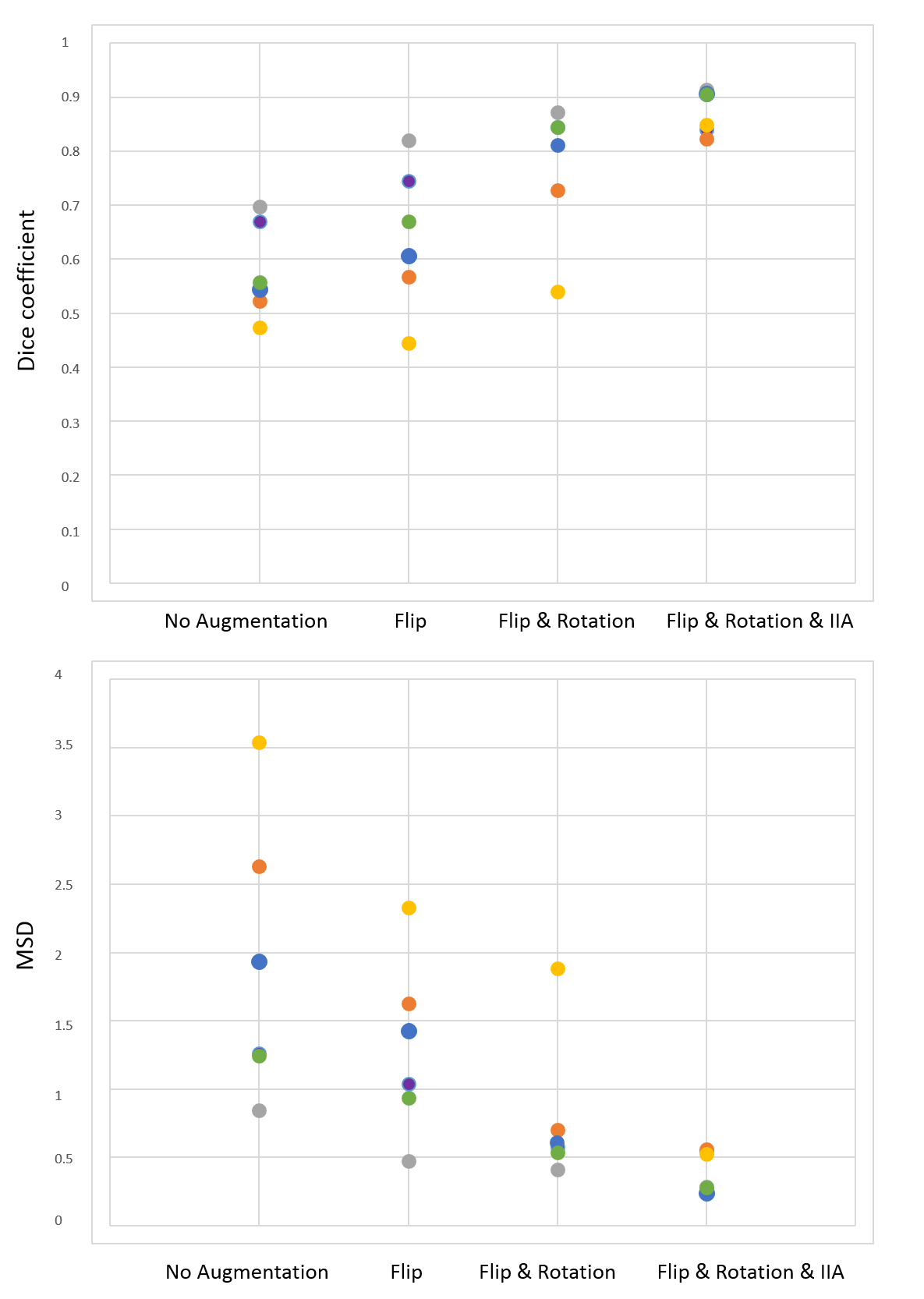}
    \caption{Performance of automatic fetal brain tissue segmentation into seven tissue classes when the network is trained without any augmentation, with flipping slices as augmentation, with flipping and rotating slices as augmentation, and with flipping, rotating and IIA. The results are expressed as the mean Dice coefficient and the mean surface distance (MSD) in mm. Each marker corresponds to the mean performance across all tissue classes in one of the six test scans.}
    \label{fig:results_augmentations}
\end{figure}

The achieved average DC and MSD for each scan in the test set are shown in Figure ~\ref{fig:results_augmentations}. The performance improved in all scans the more data augmentation was used and especially further improved when IIA was added in addition to the standard data augmentation techniques. IIA largely reduced performance differences between different scans, which standard augmentation techniques were not able to achieve. The performance improvement was particularly large in a scan in which nearly all slices showed intensity inhomogeneity (yellow marker in Figure~\ref{fig:results_augmentations}). With IIA, the segmentation performance reached an accuracy comparable to that achieved on scans with fewer artifacts.

\begin{figure}[ht!]
    \includegraphics[width=0.9\textwidth]{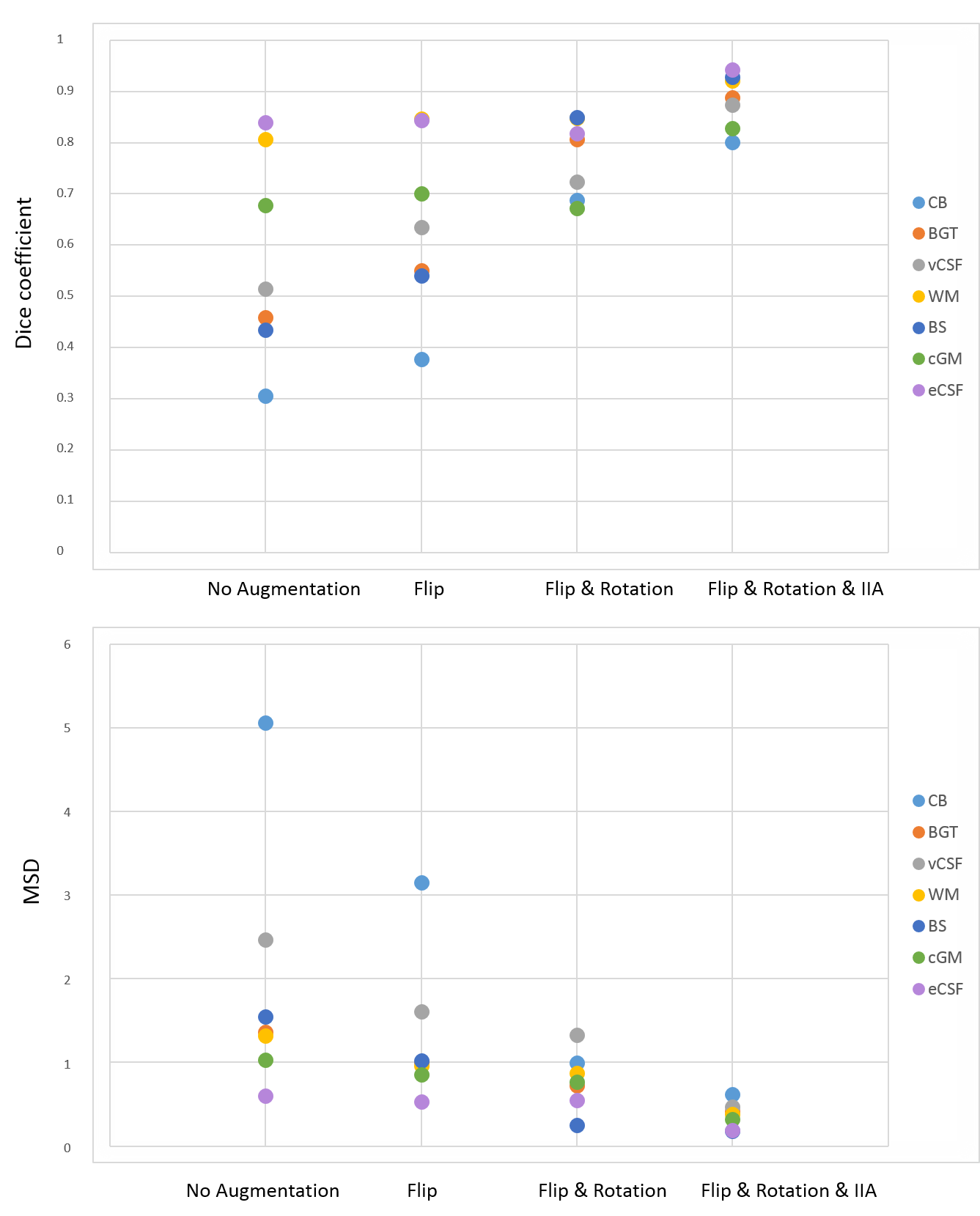}

    \caption{Performance of automatic fetal brain tissue segmentation for each of the seven tissue classes when the network is trained without any augmentation, with flipping augmentation, with flipping and rotating augmentation, and with flipping, rotating and IIA. The results are expressed as the mean Dice coefficient and the mean surface distance (MSD) in mm. Each marker corresponds to the mean performance across all six test scans for one of the seven tissue classes.}
    \label{fig:results_augmentations_7tissue}
\end{figure}

Additionally, the achieved average DC and MSD for each tissue class are shown in Figure ~\ref{fig:results_augmentations_7tissue}. Overall, all augmentation methods improved segmentation performance. Adding augmentation based on random flipping and rotation of the image slices improved the segmentation of CB, BS, vCSF, BGT noticeably but not the segmentation of WM, cGM and eCSF. IIA was able to further improve the segmentation performance in all tissue classes including WM, cGM and eCSF. Overall, CB segmentation performance benefited the most from all augmentations where the DC increased from 0.3 to 0.8. Furthermore, augmentation based on random rotation severly influenced performance in segmentation of BGT, BS and CB.

\begin{table}[ht!]
    \resizebox{\textwidth}{!}{
        \begin{tabular}{c|n{1}{2}n{1}{1}n{1}{1}n{1}{1}n{1}{3}n{1}{3}n{1}{1}n{1}{1}n{1}{1}}
        & {CB} & {BGT} & {vCSF} & {WM} & {BS} & {cGM} & {eCSF} \\ \hline
        Proposed Method &0,793820412	&0,931386794	&0,874061628&	0,919396137&	0,946435558	&0,834999148	&0,944196498	\\
        Hebas et al. \cite{habas2010atlas} &{{-}}&	{{-}}&	0,90	&0,90	&{{-}}	&0,82	&{{-}}	\\
        Serag et al. \cite{serag2012multi}&{{-}}&{{-}}& 0,92&0,90&{{-}}& 0,84&{{-}}\\
        \end{tabular}
    }
    \caption{Segmentation performance of the proposed method and of other methods evaluated with Dice coefficient. Performance of previous methods is taken from the literature. Hence, this comparison can be used as indication only.}
    \label{tab:compare}
\end{table}
\subsection{Impact of extent of intensity inhomogeneity augmentation}
\begin{figure}[ht!]
    \includegraphics[width=\textwidth]{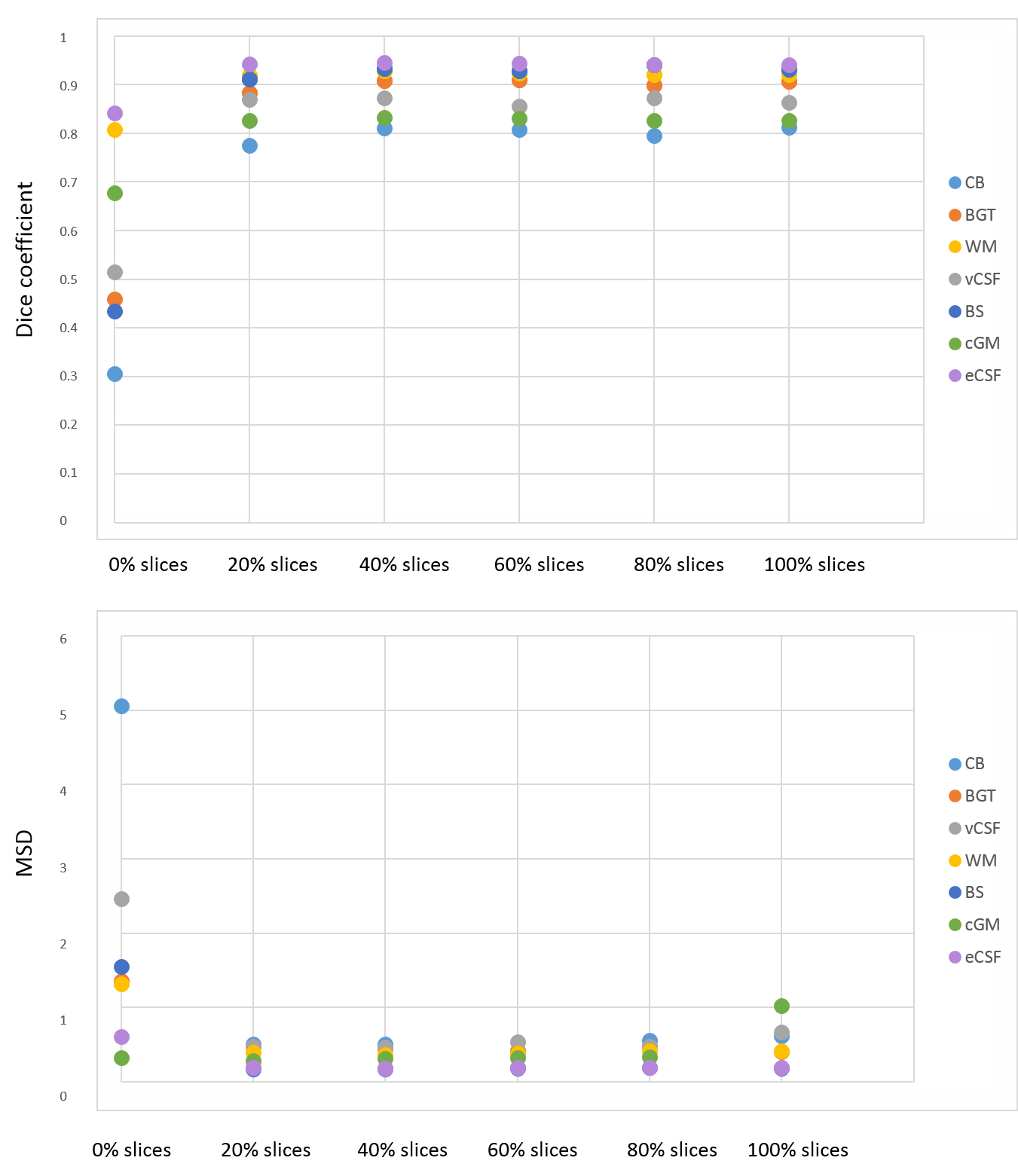}
    \caption{Performance of fetal brain tissue segmentation into seven tissue classes with different proportion of slices with synthetic intensity inhomogeneity in each training batch. The results are expressed as the mean Dice coefficient (DC) and the mean surface distance (MSD) in mm. Each marker corresponds to the mean performance across all six test scans for a different tissue class.}
    \label{fig:proportoion}
\end{figure}

To assess the impact of the proportion of slices in each training batch with simulated intensity inhomogeneity on the network performance, we trained the network with different percentage of artifact free slices to slices with simulated intensity inhomogeneity. In this experiment, all slices with intensity inhomogeneity were removed from the training set \revised{(Set 1)}. We varied number of slices with added synthetic intensity inhomogeneity from 0\% to 100\%. Figure \ref{fig:proportoion} shows the obtained results. Training with only simulated slices led to slightly worse performance compared with a mix of original and manipulated slices. Even having only 20\% of the slices with synthetic intensity inhomogeneity already improved the performance substantially. There was, however, no marked improvement for larger percentages of slices with synthetic intensity inhomogeneity, the performance was comparable for percentage of 20\% to 100\%.

\subsection{Evaluation of IIA on neonatal brain segmentation}

\begin{figure}[ht!]
    \includegraphics[width=0.83\textwidth]{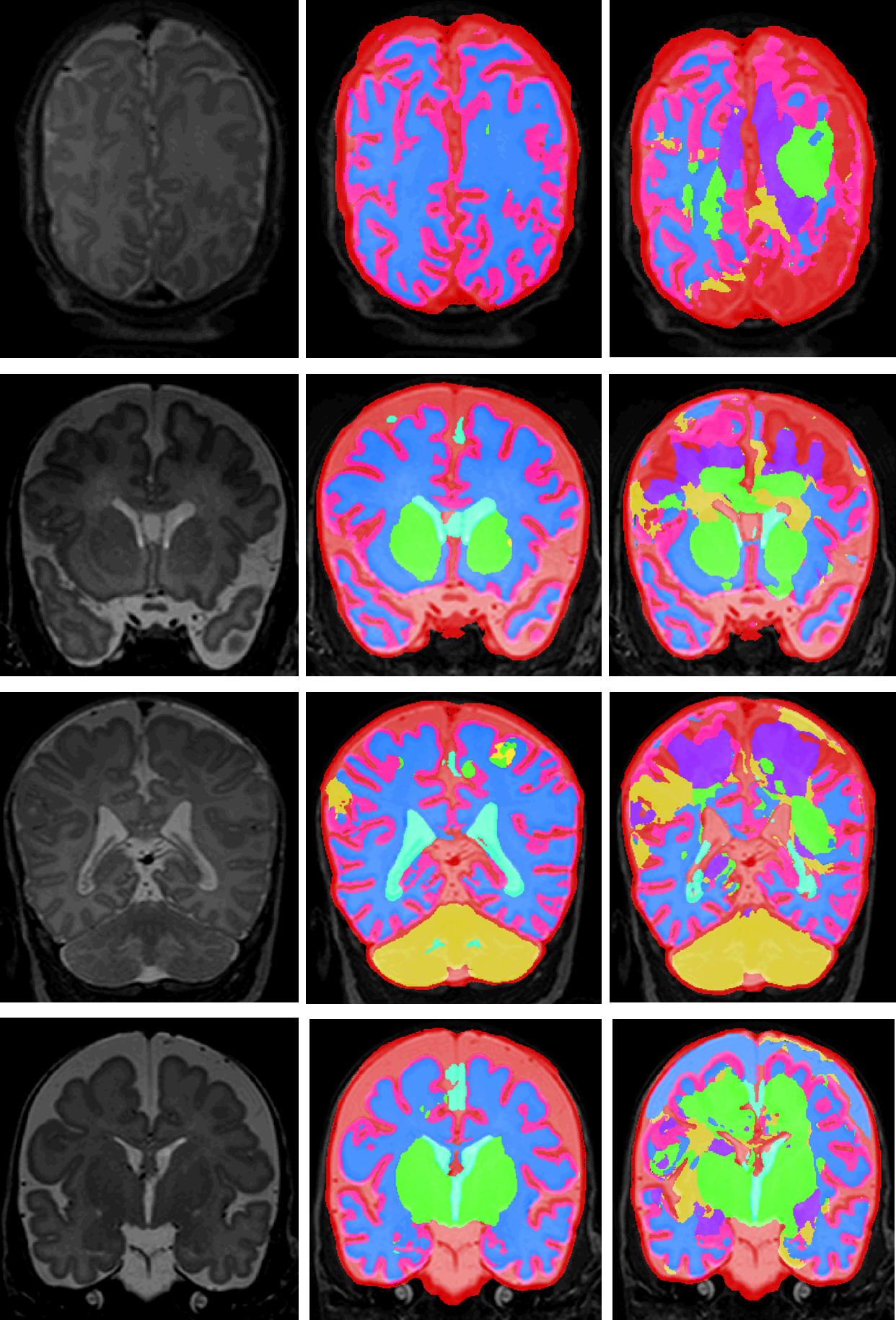}
    \begin{minipage}{0.9\textwidth}
        \threesubcaptions{Neonatal brain MRI}{With IIA}{Without IIA}
        \includegraphics[width=0.85\textwidth]{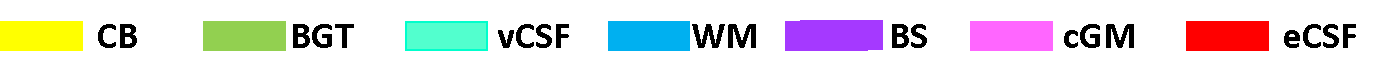}
    \end{minipage}
    \caption{Examples of brain tissue segmentation in neonatal MRI with intensity inhomogeneity artifacts. A slice from T2-weighted fetal MRI scan (first column); segmentation obtained with network using rotation, flipping and IIA (second column); segmentation obtained with network only using flipping and rotation of the slices as training data augmentation (third column)}
    \label{fig:neonate_artifact}
\end{figure}

Intensity inhomogeneity is an artifact that occurs in various of MRI scans, albeit often to lesser extent than in fetal MRI. To asses whether the proposed augmentation technique-IIA-is also able to improve the performance of a segmentation task in other MR images, we trained the brain tissue segmentation network to perform segmentation in neonatal brain MRI scans. Following previous work \cite{moeskops2015automatic} to limit the number of voxels considered in the classification, brain masks were generated with BET \cite{smith2002fast}. The network was trained with three scans and tested with the remaining two scans \revised{(Set 3)}. Like in the previous experiments, the network was first trained using standard data augmentation, i.e., using random flipping and rotation of the training slices as described in Section 3, and subsequently the network was trained additionally with IIA. The obtained segmentation results are listed in Table~\ref{tab:neonat}. 
As the two scans in the test set did not show any intensity inhomogeneity artifacts, we additionally qualitatively evaluated the segmentation performance on four scans with visible intensity inhomogeneity artifacts for which manual reference segmentations were not available. We illustrate segmentation results in the four scans without reference standard in Figure~\ref{fig:neonate_artifact}. In \cite{moeskops2015automatic} these scans were not analyzed due to the presence of artifacts. Visual inspection of the results in these scans reveals that the segmentation was more accurate when IIA was used, particularly in BGT, BS, vCSF and eCSF. 

\begin{table}[ht!]
\resizebox{\textwidth}{!}{%
\begin{tabular}{c|c|c|n{3}{3}n{1}{1}n{1}{1}n{1}{1}n{1}{3}n{1}{3}n{1}{1}|n{1}{1}}
&&& {CB} & {BGT} & {vCSF} & {WM} & {BS} & {cGM} & {eCSF} & {Mean}\\ \hline
All test slices& With IIA&DC
&0,857224
&0,883034
&0,7756945
&0,8249265
&0,764997
&0,4997655
&0,599743
&0,7436263571
\\
&&MSD
&0,918794
&0,6765545
&0,7508925
&0,384828
&0,5101205
&0,375138
&0,6250645
&0,6430919375
\\
&Without IIA &DC 
&0,857873
&0,854176
&0,737331
&0,819091
&0,7755455
&0,509778
&0,595553
&0,7356210714\\
&&MSD 
&1,464505
&1,35862
&0,617052
&0,39324
&1,383005
&0,372276
&0,6349215
&0,8890885\\
\end{tabular}
}
\caption{Performance of neonatal brain segmentation into seven tissue classes. The segmentation performance with IIA is compared with the performance of the same network without IIA. The results are expressed as the mean Dice coefficient (DC) and the mean surface distance (MSD) in mm.}
\label{tab:neonat}
\end{table}

\section{Discussion}

We presented a pipeline for automatic segmentation of the fetal brain into seven tissue classes in MRI. The method consists of two fully convolutional networks with identical U-net architectures. The first network extracts ICV and the second network performs segmentation of the brain into seven tissue classes. The results demonstrate that segmentation using the proposed data augmentation with simulated intensity inhomogeneity artifacts leads to accurate segmentations of the brain tissue classes. Moreover, we demonstrated that the method performs accurate segmentation while trained using manual reference segmentation only in slices without artifacts that occur during image acquisition. In other words, we showed that the proposed data augmentation is able to compensate for the lack of training data in which performing manual annotations is cumbersome.

Using the proposed data augmentation technique, we were able to achieve state-of-the-art segmentation performance with a substantially lower number of training scans. Our method was trained with only 6 fetal scans while previous methods used 20 \cite{gholipour2012multi} and up to 80 fetal scans \cite{serag2016accurate}. Given that manual annotation of a fetal brain MR scan into 7 tissue classes requires about 40 hours, reducing the number of the  manually annotated training scans  substantially reduces the required manual annotation effort and associated costs. 

Intensity inhomogeneity is a frequently occurring artifact in MRI and often hampers automatic image analysis due to diminishing contrast between different tissues. We demonstrated that the described method based on convolutional neural networks can become more robust to these artifacts by training with data augmented with simulated random intensity inhomogeneities. This can potentially replace or complement prepossessing steps, such as bias field corrections or volumetric reconstructions that would require acquisition of additional MR data. Simulating artifacts instead of manually annotating compromised data for training supervised methods is beneficial as manual reference annotations can be obtained more easily and with higher accuracy for artifact-free data.

Standardly used data augmentation techniques, random flipping and rotation of the image slices, improved the segmentation performance in CB, BS, BGT, and vCSF considerably but had little impact on other tissue types. Introducing training data with simulated intensity inhomogeneity further improved the segmentation performance in all tissue classes, including WM and cGM, which are often challenging to separate due to low inter-tissue contrast when intensity inhomogeneity artifacts are present. Moreover, a frequent mistake of the automatic segmentation method when training without IIA was mistaking vCSF for eCSF and vice versa. Using IIA helped to overcome this issue in many cases, presumably by forcing the network not to focus on the intensity values only but additionally on other intensity invariant information such as shape and context.


Moreover, evaluating augmentation techniques per scan showed that the MSD reduces with adding random flipping and rotating augmentation in all scans and DC improves in all scans except one (indicated with a yellow marker in Figure \ref{fig:results_augmentations}). Retrospective visual inspection revealed that this scan has intensity inhomogeneity in nearly all slices. Even though the intensity inhomogeneity is not severe in all slices, the automatic segmentation was still severely affected. Experiments show that adding IIA in the training helped to overcome this issue and increased the segmentation performance in all scans.

Furthermore, our experiments illustrate that training the network with IIA increases the segmentation performance even in slices without visible intensity inhomogeneity. IIA makes the network more robust to intensity variations in MRI, forcing the network not to focus only on the tissue intensity for assigning a label.

In the current study, the segmentation method was evaluated on fetal brain MRI acquired in the coronal plane. Since the presented method is entirely supervised it can be readily applied to fetal MRI acquired in axial or sagittal plane if manually annotated training data is available.

We relied on the U-net architecture for ICV and brain tissue segmentation in fetal MRI. However, the proposed IIA can be used for data augmentation regardless of the network architecture or even with supervised methods not based on convolutional neural networks. We did not evaluate IIA with other architectures but it would likely improve the segmentation performance of other networks with different architecture as supervised CNNs regularly profit from large and diverse training data.

Additionally, we evaluated IIA on neonatal MRI. The visual inspection shows a substantial improvement on slices with artifacts. In the images without visible artifacts the quantitative results showed slight improvement when IIA is applied. The performance of the segmentation in the neonatal brain scans is lower than obtained by our previous method \cite{moeskops2015automatic} as here presented network was not specifically adjusted for segmentation of neonatal brain tissues. However, the results clearly demonstrate the benefit of training with IIA. IIA could be readily applied to the segmentation method presented in \cite{moeskops2015automatic} enabling the multi-scale CNN to segment neonatal brain scans with intensity inhomogeneity artifacts as shown in Figure \ref{fig:neonate_artifact}.

In this study, 2D analysis was applied since fetal MR images has inter-slice motion due to the 2D MR acquisition. Additionally, in a few slices severe motion artifacts occurred. These slices were excluded from the training and test set as manually segmenting them for evaluation would be hardly feasible. Generating such slices that are heavily affected by motion artifacts in fetal MRI could be an interesting direction for future work.

\revised{We have trained the proposed method with representative data, i.e. in a supervised manner. Training with non-representative, in addition to the representative data, using transfer learning would allow increasing the training sample size.  This could be addressed in future work and it could potentially further improve segmentation performance.}

\section{Conclusion}
We presented an automatic method for brain tissue segmentation in fetal MRI into seven tissue classes using convolutional neural networks. We demonstrated that the proposed method learns to cope with intensity inhomogeneity artifacts by augmenting the training data with synthesized intensity inhomogeneity artifacts. This can potentially replace or complement preprocessing steps, such as bias field corrections, and help to substantially improve the segmentation performance.

\section*{Acknowledgment}
This study was sponsored by the Research Program Specialized Nutrition of the Utrecht Center for Food and Health, through a subsidy from the Dutch Ministry of Economic Affairs, the Utrecht Province and the Municipality of Utrecht. Furthermore, we thank Nienke Heuvelink for her help with creating the manual fetal brain segmentations.

\section*{References}
\bibliography{strings}

\end{document}